\documentclass[%
 reprint,
 amsmath,amssymb,
 aps,
]{revtex4-2}
\usepackage{graphicx}
\usepackage{dcolumn}
\usepackage{bm}

\usepackage[utf8]{inputenc}

\usepackage{siunitx} 

\usepackage{layout} 

\usepackage{hyperref}
\hypersetup{
      filecolor=blue,
      urlcolor=blue
      colorlinks=true,
      linkcolor=blue,
      citecolor=blue}

\usepackage{graphicx}

\usepackage{physics}
\usepackage{amssymb}
\usepackage{amsmath}

\usepackage[dvipsnames]{xcolor}
\usepackage{tikz}
\usetikzlibrary{3d}
\usepackage{circuitikz}
\usepackage{cuted}
\usepackage{soul}
\usepackage{mathtools}
\usepackage{lineno}

\begin{document}
\preprint{APS/123-QED}

\title{Dynamics of hole singlet triplet qubits with large g-factor differences}%
\date{\today}

\author{Daniel Jirovec$^1$}
\email{daniel.jirovec@ist.ac.at}
\author{Philipp M. Mutter$^2$}%
\email{philipp.mutter@uni-konstanz.de}
\author{Andrea Hofmann$^{1,3}$}
\author{Josip Kukucka$^1$}
\author{Alessandro Crippa$^{1,4}$}
\author{Frederico Martins$^{1,5}$}
\author{Andrea Ballabio$^6$}
\author{Daniel Chrastina$^6$}
\author{Giovanni Isella$^6$}
\author{Guido Burkard$^2$}
\author{Georgios Katsaros$^1$}

\affiliation{
$^1$Institute of Science and Technology Austria, Am Campus 1, 3400 Klosterneuburg, Austria\\
$^2$Department of Physics, University of Konstanz, D-78457 Konstanz, Germany \\
$^3$Department of Physics, University of Basel, Klingelbergstrasse 82, CH-4056 Basel, Switzerland\\
$^4$NEST, Istituto Nanoscienze-CNR and Scuola Normale Superiore, I-56127 Pisa, Italy\\
$^5$Hitachi Cambridge Laboratory, J.J. Thomson Avenue, Cambridge CB3 0HE, United Kingdom\\
$^6$L-NESS, Physics Department, Politecnico di Milano, via Anzani 42, 22100, Como, Italy\\
}

\begin{abstract}
The spin-orbit interaction is the key element for electrically tunable spin qubits. Here we probe the effect of cubic Rashba spin-orbit interaction on mixing of the spin states by investigating singlet-triplet oscillations in a planar Ge hole double quantum dot. By varying the magnetic field direction we find an intriguing transformation of the funnel into a butterfly-shaped pattern.  Landau-Zener sweeps  disentangle the Zeeman mixing effect from the spin-orbit induced coupling and show that large singlet-triplet avoided crossings do not imply a strong spin-orbit interaction. Our work emphasizes the need for a complete knowledge of the energy landscape when working with hole spin qubits. 
\end{abstract}
\maketitle
The spin-orbit interaction (SOI) allows electrical manipulation of individual spins and has therefore become a key ingredient for the realization of fully electrically controlled spin qubits \cite{Golovach2006, Bulaev2007}. In Si, for electrons it is rather weak and synthetically boosted by means of micromagnets \cite{Wu2014, Yoneda2017}. For holes, on the other hand, it is an intrinsic property which allows to perform electron dipole spin resonance (EDSR) measurements \cite{Golovach2006, Bulaev2007, Maurand2016, Crippa2018, Watzinger2018, Hendrickx2020a, Hendrickx2020}. In Ge it is particularly strong leading to Rabi frequencies beyond 100~MHz \cite{Watzinger2018, Froning2021, Wang2020}. SOI for holes can be linear or cubic in the wave vector k, with nanowire qubits favoring the former type while planar qubits the latter \cite{Bosco2021, Wang2021}. 
The SOI is not only important for single spin but also for singlet-triplet qubits as it causes an intrinsic mixing between the heavy hole (HH) and light hole (LH) bands and thereby locally affects the g-factors of the individual spins allowing to drive $S-T_0$ oscillations \cite{Jirovec2021}. In combination with an extrinsic Rashba type SOI caused by the structural inversion asymmetry induced by the heterostructure, it also mixes the $S$ and $T_-$ states contributing therefore to the observed avoided crossing. This singlet-triplet splitting $\Delta_{ST_-}$ has been extensively studied in GaAs structures. Different regimes dominated either by the SOI or the hyperfine interaction have been investigated and optimized for dynamical nuclear polarization \cite{Dickel2015, Nichol2015}. In Si, where the hyperfine contribution is much weaker \cite{Maune2012}, the relatively weak SOI is found to be the decisive factor for the size of the anticrossing \cite{HarveyCollard2019, Tanttu2019}.\\

Here we study a double quantum dot singlet triplet hole qubit realized in planar Ge. We characterize the complete spectrum of our system by observing the dynamics under different pulsing schemes and magnetic field directions. We investigate the $S-T_-$ avoided-crossing by means of Landau-Zener tunneling  and find that a typical cubic SOI parametrized by an in-plane spin-flip tunneling term $t_{SO}$ is insufficient to describe the observed angular B-field dependence. In fact, the different g-factor anisotropy in the two quantum dots greatly enhances $\Delta_{ST_-}$ in the in-plane magnetic field direction and influences the system dynamics.

\begin{figure*}
    \centering
    \includegraphics[width = 
    \textwidth]{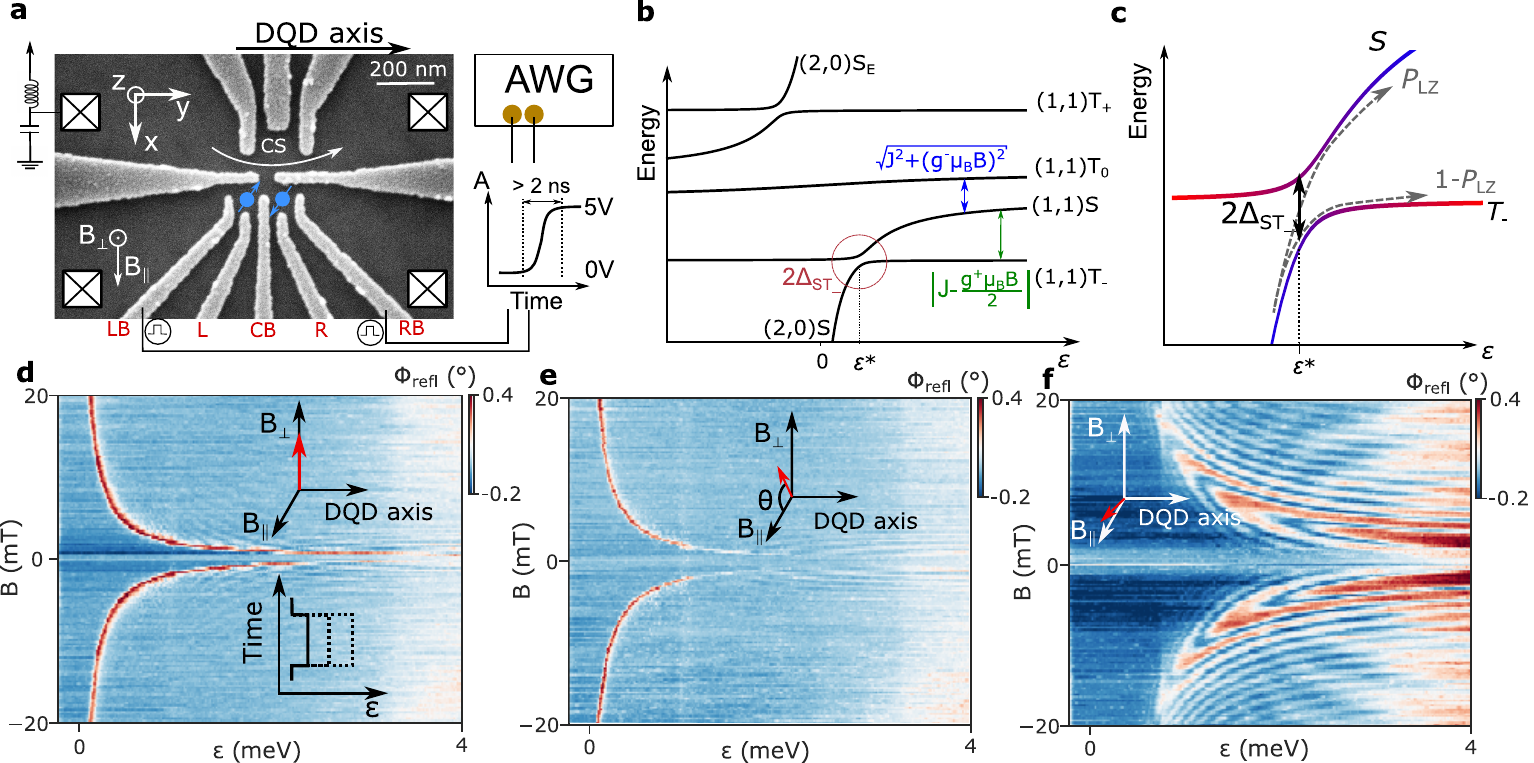}
    \caption{a) SEM image of the device consisting of a double dot electrostatically defined by gates LB, L, CB, R and RB. The charge sensor (CS) is coupled to an LC-resonator for Ohmic-reflectometry via a lock-in amplifier. Pulses are applied to gates LB and RB via an arbitrary waveform generator with a minimum rise time of 2 ns (inset).
    b) The energy level diagram as a function of detuning  highlights the relevant energy splittings between $S$ and $T_0$ and $S$ and $T_-$. At $\epsilon = \epsilon^*$, $S$ and $T_-$ anticross with a splitting $2 \Delta_{ST_-}$. 
    c) Energy level diagram of the states involved in the passage over the avoided crossing (red circle in b)). The probability $P_{LZ}$ to maintain the initial state after a single passage over the avoided crossing is given by the Landau-Zener formula.
    d) Pulsing $\epsilon$ over the $S-T_-$ degeneracy will result in a mixing of the two states when $\epsilon = \epsilon^*$. Here we show the resulting return signal for a magnetic field applied at an angle $\theta=\SI{90}{\degree}$ from the in-plane direction. A high signal corresponds to a larger triplet return probability. The lower inset displays the pulse sequence where only the amplitude $\epsilon$ is varied. e) At $\theta=\SI{60}{\degree}$ compared to d) the degeneracy is sharper indicating a smaller value of $\Delta_{ST_-}$. f) For  $\theta=\SI{10}{\degree}$ the degeneracy is not a funnel-like sharp line but rather an oscillation pattern resembling a butterfly shape. Such features appear when the $S-T_-$ mixing term becomes substantial. At low $\epsilon$ $S-T_-$-like oscillations are prominent. At higher detuning also $S-T_0$-like oscillations become more visible.}
    \label{fig:Funnel}
\end{figure*}

A scanning electron microscope (SEM) image of the device under consideration is depicted in Fig.~\ref{fig:Funnel}a and further details can be found in \cite{Jirovec2021}. A hole gas confined in a Ge quantum well is buried $\SI{20}{\nano\meter}$ below the surface. A charge sensor (CS) connected to a radio-frequency (rf) reflectometry circuit is used to read out the charge state of the gate defined DQD. For qubit state selective read-out we rely on Pauli spin blockade. Fast detuning pulses are applied to gates LB and RB with an arbitrary waveform generator (AWG) which has a pulse-rise time of $\tau_{rise} \approx \SI{2}{\nano\second}$ (inset of Fig. \ref{fig:Funnel}a). Throughout this work we apply a small magnetic field in a plane perpendicular to the quantum dot axis, $\textbf{B} = (B\cos(\theta), 0, B\sin(\theta))$, where $\theta$ describes the tilt angle from the in-plane direction. We tune the DQD to an effective charge transition (2,0)$\leftrightarrow$(1,1), with ($n_L$, $n_R$) where $n_L$ ($n_R$) denotes the effective hole number in the left (right) quantum dot (see Supplementary Fig. \ref{fig:StabilityDiagram}). We emphasize that the real hole number in the left dot is $n_L+2$ while in the right dot we cannot determine the exact hole number. The tunnel coupling between the dots is described by $t_C$ while the energy detuning between the $S(2,0)$ and $S(1,1)$ state is parametrized by $\epsilon$. Each quantum dot is characterized by an out-of-plane and an in-plane g-factor, $g_\perp$ and $g_\parallel$, respectively. However, the dynamics of singlet-triplet qubits is only sensitive to differences in, or the average of, the Zeeman energy between the dots, and hence we define $g^\pm = g^L \pm g^R$  as the g-factor difference and sum. The energy spectrum of the system (the complete Hamiltonian $H_{tot}$ is derived in Supplementary section \ref{sec:model}) is depicted in Fig. \ref{fig:Funnel}b as a function of $\epsilon$. At $\epsilon = \epsilon^*$ the $S$ and $T_-$ states anticross.\\
We start by mapping out the $S-T_-$ degeneracy as a function of magnetic field angle with the funnel technique \cite{Petta2005}. Here, mixing between $S$ and $T_-$ is induced by pulsing the system close to $\epsilon = \epsilon^*$. Mixing depends both on the size of the avoided crossing and the mixing time $\tau_S$. We apply a rapid pulse of duration $\tau_S = \SI{65}{\nano\second}$ and varying $\epsilon$ (inset of Fig. \ref{fig:Funnel}d). Fig.\ref{fig:Funnel}d,e and f depict the phase response of the charge sensor in the measurement point as a function of the pulse amplitude on $\epsilon$ and the magnetic field strength for $\theta = \SI{90}{\degree}$, $\SI{60}{\degree}$ and $\SI{10}{\degree}$, respectively. A high return signal corresponds to a larger triplet probability. In the out of plane direction we observe the expected funnel shape of the $S-T_-$ degeneracy. At $\SI{60}{\degree}$ we similarly observe a typical funnel shape, however, we notice the line to be fainter which indicates a smaller $\Delta_{ST_-}$. The picture drastically changes towards the in-plane direction where the $S-T_-$ degeneracy develops interference fringes with a pattern resembling a butterfly; 2 main components can be attributed to $S-T_-$ oscillations at low detuning and $S-T_0$ oscillations becoming more prominent at high detuning. The prominent $S-T_-$ oscillations are an indication of a large coupling term in the in-plane direction.
The angular anisotropy of the funnel pattern, further exemplified in Fig. \ref{fig:SuppFunnelvsAngle}, is the main focus of this work and requires knowledge of the full Hamiltonian and therefore an understanding of the interplay between the g-factor anisotropy and the spin-flip element $t_{SO}$. \\

\begin{figure}
    \centering
    \includegraphics[width = 0.5\textwidth]{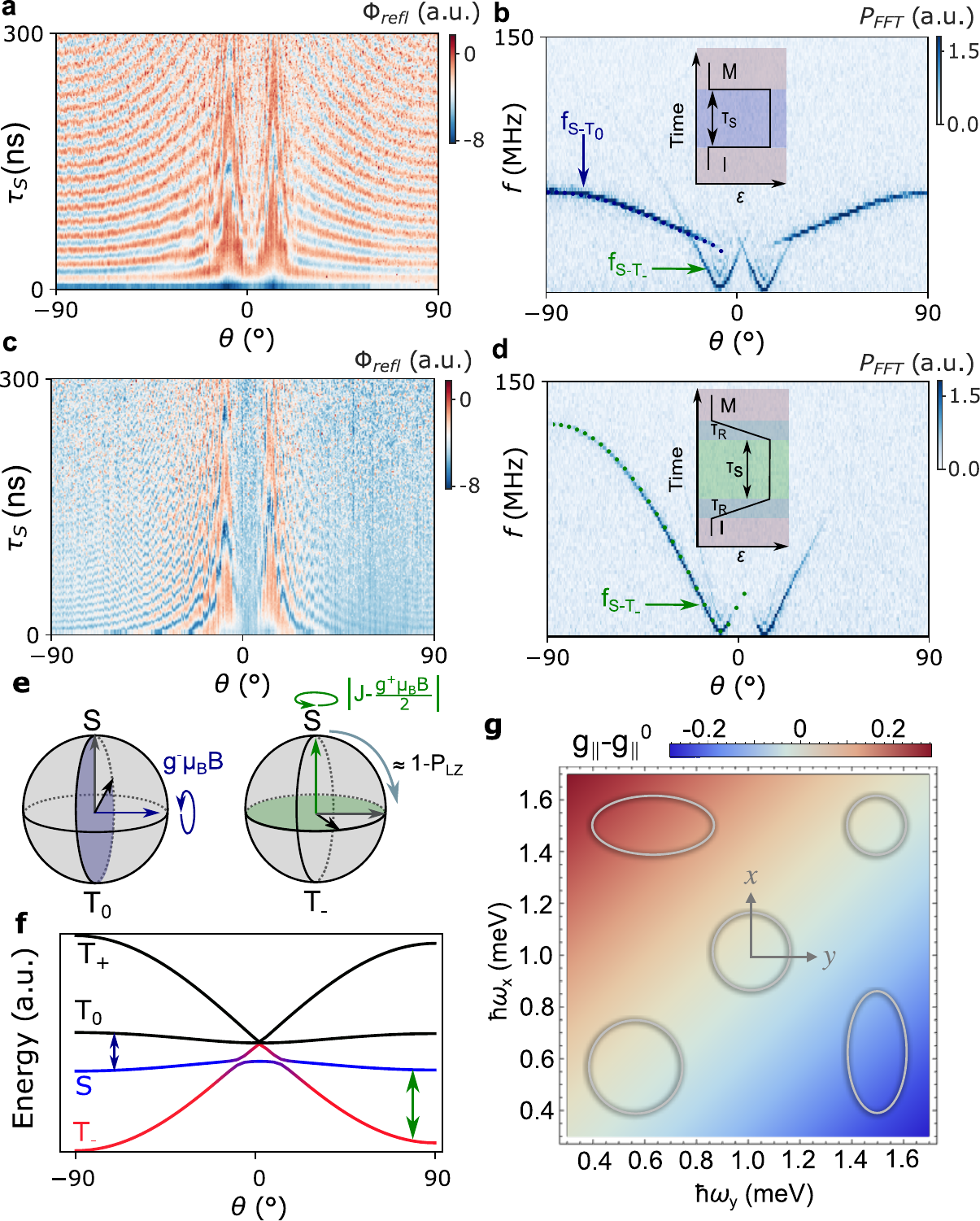}
    \caption{a) Oscillation amplitude of the singlet state in the measurement point as a function of separation time and magnetic field angle at $B = \SI{2}{\milli\tesla}$. b) FFT of a) revealing the oscillation frequency anisotropy. The blue dotted line is a fit to our model. We find a small offset of $\SI{100}{\micro\tesla}$ in the perpendicular field which leads to a small asymmetry in the FFT plots. The inset shows the pulse shape where the system is swept to $\epsilon = \SI{4}{\milli e \volt}$ in the shortest time enabled by the AWG.
    c) $S-T_-$ oscillations under the same conditions as a) but with a longer ramp time ($\tau_R = \SI{100}{\nano\second}$). d) FFT of c) showing $f_{S-T_-}$ as the green dotted fit line. The $S-T_0$ oscillations are completely suppressed by the slow ramping to high $\epsilon$ highlighted by the pulse sequence in the inset. e) Schematic of the time evolution at high $\epsilon$ for the $S-T_0$ ($S-T_-$) qubit on the left (right). The amount of mixing between $S$ and $T_-$ depends on $P_{LZ}$ (light blue arrow) and constitutes a second rotation axis for the qubit \cite{Petta2010}. f) The energy dispersion of the eigenstates of $H_{tot}$ at $\epsilon = \SI{4}{\milli e  \volt}$ as a function of $\theta$ reproduce the frequencies seen in b (d) with the blue (green) arrow highlighting the visible transition. 
    g) Effect of the confinement on the in-plane g-factors for a quantum well width of $\SI{20}{\nano\meter}$ according to Eq.~\eqref{eq:effective_in_plane_g_factor}. On top, we schematically show possible dot geometries in real space.
    }
    \label{fig:RabivsAngle}
\end{figure}

In order to extract the g-factor anisotropy we rely on singlet-triplet oscillations. After initialization in $S(2,0)$, appropriate pulses to (1,1) induce either $S-T_0$ or $S-T_-$ oscillations.  The probability to maintain the initial eigenstate of the system after a sweep with ramp time $\tau_R$ is given by the Landau-Zener formula $P_{LZ} = \exp(-\frac{2\pi\Delta_{ST_-}^2}{\hbar v})$, where $\hbar$ is the reduced Planck constant and $v =  |\frac{dE}{dt}| = |\frac{dE}{d\epsilon}\frac{d\epsilon}{dt}| = |\frac{dJ(\epsilon)}{d\epsilon}|_{\epsilon=\epsilon^*}\frac{\Delta\epsilon}{\tau_R}$ is the velocity  calculated at $\epsilon = \epsilon^*$ and $J(\epsilon) = \sqrt{\frac{\epsilon^2}{4}+2t_C^2}-\frac{\epsilon}{2}$ is the exchange energy (Fig. \ref{fig:Funnel}b and c). If $v$ satisfies the diabatic condition ($P_{LZ}\approx 1 $) $S-T_0$ oscillations with a frequency $f = \frac{1}{\hbar}\sqrt{J^2+(g^-\mu_B B)^2}$ will be favored. With $P_{LZ} < 1$ the $S-T_0$ oscillations are suppressed and the qubit is initialized in a superposition of $S$ and $T_-$. After a time $\tau_S$ the system is pulsed back to the measurement point where another non-diabatic passage will cause an interference between the two states similar to a Mach-Zhender interferometer \cite{Shevchenko2010}. The accumulated phase difference is then given by $\phi = 2\pi f_{S-T_-}\tau_S \approx \frac{\tau_S}{\hbar}|J-\frac{1}{2}g^+\mu_B B|$ \cite{Petta2010} (see Fig. \ref{fig:Funnel}b).
As the oscillation frequency of the $S-T_0$ qubit is proportional to $g^-$ while the one of the $S-T_-$ qubit is proportional to $g^+$ (Fig. \ref{fig:Funnel}b and \ref{fig:RabivsAngle}e) we can extract the individual g-factors without the need for EDSR. 
We fix the magnetic field at $|B| = \SI{2}{\milli\tesla}$ and observe the oscillations resulting from a fast pulse ($\tau_R = \tau_{rise} = \SI{2}{\nano\second}$) in Fig. \ref{fig:RabivsAngle}a and a ramped pulse with ramp time $\tau_R = \SI{100}{\nano\second}$  of amplitude $\epsilon = \SI{4}{\milli e \volt} $ and duration $\tau_S$ in Fig. \ref{fig:RabivsAngle}c as we rotate the field.
From the fast Fourier transform (FFT) in Fig. \ref{fig:RabivsAngle}b and d we extract the oscillation frequency $f_{S-T_0}$ (blue dots) and $f_{S-T_-}$ (green dots). We notice that for $\theta \in [- \SI{25}{\degree},+\SI{25}{\degree}]$ in both FFT plots the $S-T_-$ frequency is visible suggesting that a large coupling term is present at these magnetic field directions inducing a non-diabatic passage, in line with the observations in Fig. \ref{fig:Funnel}f. Moreover, in Fig. \ref{fig:RabivsAngle}d the FFT power vanishes for $\theta\approx\SI{60}{\degree}$ indicating that the ramp time $\tau_R$ induces a completely diabatic passage over the avoided crossing. This is in line with Fig. \ref{fig:Funnel}e where we observed a sharper $S-T_-$ degeneracy characteristic of a smaller mixing term. 

The lines arising in the FFT plots can be fit by the energy splitting between the three lowest lying states of the system depicted in Fig. \ref{fig:RabivsAngle}f with $ g_\perp^+ =12.00$, $g_\perp^-=2.04$, $g_\parallel^+ = 0.10$ and $g_\parallel^- = 0.43$ and $t_C = \SI{11.38}{\micro e \volt}$. The latter is extracted from exchange oscillation measurements (see Supplementary Fig. \ref{fig:Exchange}).
By further fitting our model to the observed dynamics of the system under fast pulsing for magnetic field angles close to the in-plane direction (see Supplementary Sec. \ref{sec:SmallAngles}) we confirm the extracted values and find good agreement between the observed frequency lines and our predictions.

Interestingly $|g_\parallel^-| > |g_\parallel^+|$ while $|g_\perp^-|<|g_\perp^+|$. This means that the g-factors in the out of plane direction have the same sign while they exhibit opposite signs in the in plane direction. To understand this observation we investigate the effect of the dot geometry on the g-factors, anticipating a particularly accentuated effect on the in-plane g-factor due to its small zeroth order value of $g_{\parallel}^0 \sim 0.2$. This value is considered to be phenomenological in the sense that it incorporates corrections due to system specific influences such as strain and material composition~\cite{Marie1999, Winkler2003, Trifonov2021}. As is shown in Supplementary Sec.~\ref{sec:DotGeo} by using the semi microscopic Luttinger-Kohn Hamiltonian as a starting point, the effects of the intrinsic HH-LH mixing and an elliptical confinement potential can combine to yield g-factor renormalizations. While the correction to the out-of-plane g-factor is $\vert \delta g_{\perp} \vert < 10^{-2}$ for the values considered and hence negligible, the in-plane g-factor can be altered considerably,
	\begin{align}
	\label{eq:effective_in_plane_g_factor}
		&g_{\parallel} = g_{\parallel}^0 - \xi_1 \frac{\hbar(\omega_x - \omega_y) }{\hbar(\omega_x + \omega_y) - \xi_2 \Delta}.
	\end{align}
Here, $\xi_1 \approx 20.3 $ and $\xi_2 \approx 6.0 $ are material specific constants, $\Delta$ is the HH-LH splitting and $\hbar \omega_{x,y}$ are the in-plane confinement energies. It can be seen from Fig.~\ref{fig:RabivsAngle}g that the in-plane g-factor corrections can be negative in one dot but not in the other for opposite elliptical confinement.\\

\begin{figure}
\includegraphics[width = 0.5\textwidth]{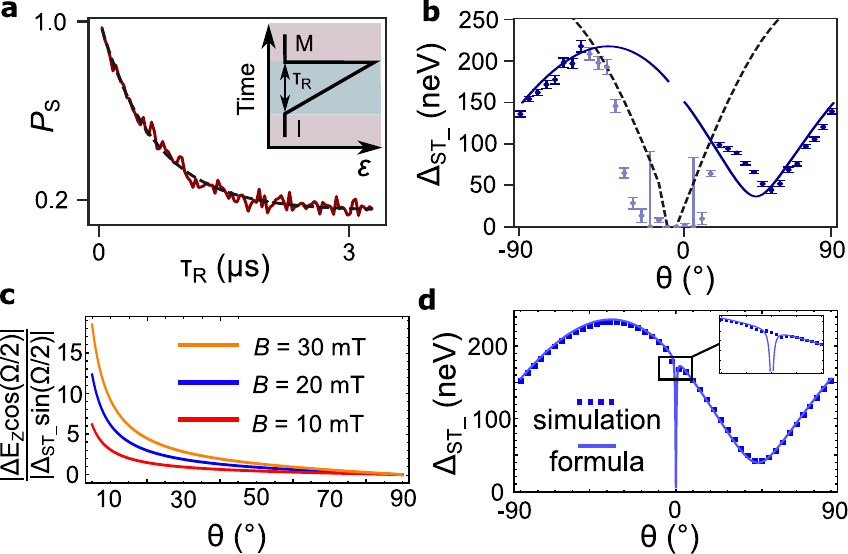}
	\caption{a) The single LZ passage pulse sequence (inset) leads to a singlet return probability $P_{S}$ that decays exponentially with the ramp time $\tau_\mathrm{R}$. A fit to the Landau-Zener transition formula (black dashed line) allows to extract $\Delta_{ST_-}$. b) $\Delta_{ST_-}$ as a function of magnetic field angle. The extracted $\Delta_{ST_-}$ is fit to Eq.~\ref{eq:Delta_ST} with $t_x$ and $t_y$ as fitting parameters (solid blue line). The black dashed line represents the maximum $\Delta_{ST_-}$ as a function of $\theta$ that can be reliably measured by a single LZ passage. The light colored data points are, therefore, excluded from the fit. c) Comparison between the two contributions to $\Delta_{ST_-}$. At small angles, the Zeeman splitting $\Delta E_Z$ can exceed the spin-orbit splitting $\Delta_{SO}$ by one order of magnitude even at $B= \SI{20}{\milli\tesla}$. d) Comparison between the analytical result (solid line, Eq.~\eqref{eq:Delta_ST}) and numerical simulations (squares) for the $S-T_-$ splitting $\Delta_{ST_-}$. We find excellent agreement except for a narrow region around $\theta=0$ (inset), where the analytical expression fails due to the small in-plane Zeeman energies. We use the values of the g-factors and the spin-orbit vector extracted in the text in panels c) and d).}
	\label{fig:TheoryFig}
\end{figure}

We now turn to extract $t_{SO}$ by analyzing $\Delta_{ST_-}$ in more detail. After calibrating the sweep rate and position of the $S-T_-$ degeneracy (Supplemetary Fig. \ref{fig:SuppFunnelvsAngle}), we perform Landau-Zener sweeps at $|B|= \SI{20}{\milli\tesla}$ and extract $\Delta_{ST_-}$ from $P_{LZ}$ (Fig. \ref{fig:TheoryFig}a). We vary $\tau_R$ during the first passage over the avoided crossing, creating a superposition of $S$ and $T_-$, and keep the return sweep diabatic in order to maintain this superposition (inset of Fig. \ref{fig:TheoryFig}a). The extracted $\Delta_{ST_-}$ is reported in Fig. \ref{fig:TheoryFig}b. In general, $\Delta_{ST_-}$ may depend on effects influencing the hole spins such as the g-factor differences in the two dots, the SOI and possible effective magnetic field gradients caused by the hyperfine interaction \cite{Stepanenko2012}. While the hyperfine interaction can result in a strong out-of-plane hyperfine component $\delta b_Z$ for HH states due to a special Ising-type form \cite{Fischer2008}, the inhomogeneous dephasing times extracted for $B_\perp$ of $\approx \SI{700}{\nano\second}$ at $\SI{1}{\milli\tesla}$ in Ref.~\cite{Jirovec2021} give an upper limit for the hyperfine component $\delta b_Z < \SI{2}{\nano e \volt}$, suggesting that the effects of the nuclear spin bath may safely be neglected.\\ In planar HH DQD systems the SOI can be parametrized by a real in-plane spin-orbit vector $\mathbf{t}_{SO} = (t_x, t_y, 0)$. Such in-plane spin-flip tunneling terms stem from the cubic Rashba SOI~\cite{Mutter2021}, while this type of SOI does not induce out-of-plane terms $t_z$. In a basis in which the total Hamiltonian is diagonal in the absence of the SOI and g-factor differences, the $S-T_-$ splitting has the form~\cite{Mutter2021ST}
	\begin{align}
	\label{eq:Delta_ST}
		\Delta_{ST_-} = \left\vert \Delta_{SO} \sin \left( \frac{\Omega }{2} \right)  + \Delta E_Z \cos \left( \frac{\Omega }{2} \right)  \right\vert,
	\end{align}
where the spin-orbit splitting $\Delta_{SO} $ and the Zeeman splitting $\Delta E_Z$ due to anisotropic site-dependent g-tensors read
	\begin{align}
		& \Delta_{SO} = t_y + i t_x \frac{g_{\perp}^+ \sin \theta}{\sqrt{( g_{\parallel}^+ \cos \theta)^2 + (g_{\perp}^+ \sin \theta)^2}}, \\
		& \Delta E_Z = \frac{\mu_B B}{4 \sqrt{2}} \frac{( g_{\parallel}^- g_{\perp}^+ - g_{\parallel}^+ g_{\perp}^- ) \sin (2 \theta)}{\sqrt{( g_{\parallel}^+ \cos \theta)^2 + (g_{\perp}^+ \sin \theta)^2}},
	\end{align}
and $\Omega = \arctan(2\sqrt{2}t_C/\epsilon^*)$ is the mixing angle at the anticrossing. The analytical result~\eqref{eq:Delta_ST} agrees well with the numerical results obtained by exact diagonalization of the system Hamiltonian for all $\theta$ except in  a narrow region around $\theta=0$ ($\vert \theta \vert \lesssim 2^{\circ}$, Fig.~\ref{fig:TheoryFig}d). We attribute these deviations to the small in-plane Zeeman energies which violate the assumption of an isolated two-level system made when deriving~\eqref{eq:Delta_ST}  (see Supplementary Section~\ref{sec:model}). Due to the opposite sign g-factor corrections in the dots the Zeeman splitting $\Delta E_Z$ can be the dominant contribution to $\Delta_{ST_-}$, exceeding the spin-orbit splitting by one order of magnitude at small angles. Even when the magnetic field has a large out-of-plane component, the effect of different g-factors can contribute crucially to $\Delta_{ST_-}$ (Fig.~\ref{fig:TheoryFig}c).

\begin{figure}
    \centering
    \includegraphics[width= 0.5\textwidth]{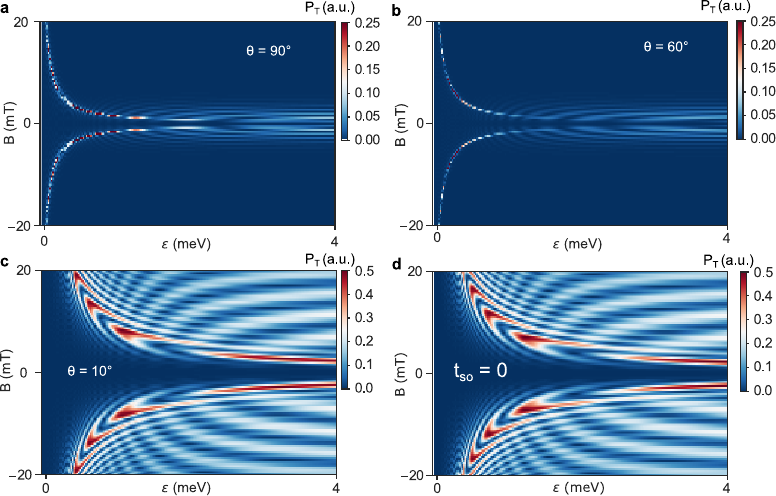}
    \caption{Funnel simulations with the master equation approach using qutip. a) Funnel for $\theta = \SI{90}{\degree}$ showing the $S-T_-$ degeneracy as an increased triplet return probability. For large $\epsilon$ $S-T_0$ oscillations are visible but decay at higher fields due to relaxation processes \cite{Barthel2012}. b) Funnel for $\theta = \SI{60}{\degree}$. Here the $S-T_-$ degeneracy is fainter than in a) because the interplay $t_{SO}$ and $g_\parallel^-$ reduces the $S-T_-$ splitting. c) Funnel for $\theta = \SI{10}{\degree}$ which now resembles a butterfly. The $S-T_-$ degeneracy evolves into an obvious oscillation pattern which interferes at large $\epsilon$ with the $S-T_0$ oscillations. The simulations take the model Hamiltonian with the experimentally extracted parameters as input and perform the time evolution calculation returning the combined triplet return probability ($(P_{T_+}+P_{T_-}+P_{T_0})$). All the simulations nicely reproduce the experimental data observed in Fig. \ref{fig:Funnel}. d) A simulation with $t_{SO} = 0$ but with all the other values of the model as in c) again reveals the butterfly shape of the $S-T_-$ degeneracy. Now even more oscillations can be seen as the $t_{SO}$ term does not reduce the size of $\Delta_{ST_-}$ caused by the g-factor difference at $\theta = \SI{10}{\degree}$.} 
    
    \label{fig:FunnelSimulation}
\end{figure}
The extracted $\Delta_{ST_-}$ in Fig. \ref{fig:TheoryFig}b can be fit by the model with $t_x$ and $t_y$ as free parameters and $t_C$, $g_\perp^+$,$g_\perp^-$,$g_\parallel^+$,$g_\parallel^-$ extracted from previous measurements. Between -25 and $\SI{+25}{\degree}$ the splitting seems to drop to zero as the Landau-Zener assumptions of diabatic return sweeps are not met and an extraction of $\Delta_{ST_-}$ is not accurate. The black dashed line corresponds to the maximum $\Delta_{ST_-}$ that allows a diabatic passage with a rise time of $\SI{2}{\nano\second}$ of our pulses ($P_{LZ,max} = 0.99 = \exp(-\frac{2\pi\Delta_{ST,max}^2}{\hbar v})$). 
The model fits the dark blue datapoints with $t_{x} = 129.0 \pm \SI{18.0}{\nano e\volt}$ and $t_y = -369.8 \pm \SI{13.8}{\nano e \volt}$, yielding the total spin-flip tunneling element $t_{SO} = \sqrt{t_x^2+t_y^2} = \SI{392.0}{\nano e \volt}$. \\
Having characterized all the elements in the Hamiltonian from independent measurements we can now reproduce the Funnel measurements in Fig.\ref{fig:Funnel} (Fig. \ref{fig:FunnelSimulation}). In particular the sharper line at $\theta = \SI{60}{\degree}$ (Fig. \ref{fig:FunnelSimulation}b) as well as the $S-T_-$ oscillations for $\theta = \SI{10}{\degree}$ (Fig. \ref{fig:FunnelSimulation}c) reflect what we observe in the data. Even with $t_{SO} = 0$ the in-plane g-factor difference induces $S-T_-$ oscillations (Fig. \ref{fig:FunnelSimulation}d) further confirming its dominant role in determining the size of $\Delta_{ST_-}$.\\
Finally, we remark that in systems with a nonlinear SOI the spin-orbit length inferred from $t_{SO}$ cannot be regarded as a good figure of merit for material comparison anymore. Following Ref.~\cite{Stepanenko2012}, we express $t_{SO}$ in terms of the spin-conserving tunneling element $t_C$ using the explicit expressions obtained in Ref.~\cite{Mutter2021}. We find 
    \begin{align}
    \label{eq:t_SO_t_C}
        t_{SO} = \frac{4t_C}{3}  \frac{d}{l_{SO}},
    \end{align}
where $d$ is the interdot distance and the spin-orbit length for a cubic Rashba SOI reads
    \begin{align}
	\label{eq:Spin_orbit_length}
		l_{SO} = \frac{1}{\lambda_R} \frac{4 \hbar \omega_0 + d^2 m \omega_0^2}{\sqrt{1-S^2} d^2 m^3 \omega_0^3}
	\end{align}
for circular dots. Here, $\hbar \omega_0$ is the in-plane confinement energy, $m$ is the in-plane HH mass, $S$ is the overlap between the left and right dot states, and $\lambda_R$ is the effective cubic Rashba coefficient with units $[\lambda_R] = $ energy/momentum$^3$. One can see that it depends not only on material parameters via $\lambda_R$ but also on the dot geometry via $\omega_0$, $d$ and $S$ (see Supplementary Sec.~\ref{sec:spin-orbit_vector_field_length}). \\

In conclusion, we have reconstructed the total Hamiltonian of our system from the dynamics of singlet-triplet oscillations. We have demonstrated that the g-factor anisotropy and in particular the in-plane g-factor sign difference can lead to a considerable contribution to $\Delta_{ST_-}$ in the in-plane direction. On the other hand, due to the interplay between $t_{SO}$ and the g-factors, angles close to the out-of-plane direction are suitable for $S-T_0$ qubit operation as the main driving terms ($g_{\perp}^-\sin\theta$) are still large but the leakage to $T_-$ during initialization and read-out is suppressed (see Fig. \ref{fig:VvsTheta}). Our results emphasize that a detailed understanding of the g-factor anisotropy and the SOI are key in designing future qubit experiments in two-dimensional Ge hole gases and other systems where a strong SOI renormalizes the g-factors.

\textbf{ACKNOWLEDGMENTS}
This research was supported by the Scientific Service Units of IST Austria through resources provided by the MIBA Machine Shop and the nanofabrication facility. This
project has received funding from the European Union's
Horizon 2020 research and innovation program under the
Marie Sklodowska-Curie grant agreement No. 844511, No. 75441, and by the FWF-P 30207, I05060 and M 3032-N projects. A.B. acknowledges support from the EU Horizon-2020 FET project microSPIRE, ID: 766955. P.M.M. and G.B. acknowledge funding by the Deutsche Forschungsgemeinschaft (DFG – German Research Foundation) under project number 450396347.

\bibliography{SOIBibliography}

\begin{thebibliography}{10}
\expandafter\ifx\csname url\endcsname\relax
  \def\url#1{\texttt{#1}}\fi
\expandafter\ifx\csname urlprefix\endcsname\relax\def\urlprefix{URL }\fi
\providecommand{\bibinfo}[2]{#2}
\providecommand{\eprint}[2][]{\url{#2}}

\bibitem{Golovach2006}
\bibinfo{author}{Golovach, V.~N.}, \bibinfo{author}{Borhani, M.} \&
  \bibinfo{author}{Loss, D.}
\newblock \bibinfo{title}{Electric-dipole-induced spin resonance in quantum
  dots}.
\newblock \emph{\bibinfo{journal}{Physical Review B}}
  \textbf{\bibinfo{volume}{74}}, \bibinfo{pages}{165319}
  (\bibinfo{year}{2006}).

\bibitem{Bulaev2007}
\bibinfo{author}{Bulaev, D.~V.} \& \bibinfo{author}{Loss, D.}
\newblock \bibinfo{title}{Electric dipole spin resonance for heavy holes in
  quantum dots}.
\newblock \emph{\bibinfo{journal}{Physical Review Letters}}
  \textbf{\bibinfo{volume}{98}}, \bibinfo{pages}{097202}
  (\bibinfo{year}{2007}).

\bibitem{Wu2014}
\bibinfo{author}{Wu, X.} \emph{et~al.}
\newblock \bibinfo{title}{Two-axis control of a singlet-triplet qubit with an
  integrated micromagnet}.
\newblock \emph{\bibinfo{journal}{Proceedings of the National Academy of
  Sciences}} \textbf{\bibinfo{volume}{111}}, \bibinfo{pages}{11938--11942}
  (\bibinfo{year}{2014}).

\bibitem{Yoneda2017}
\bibinfo{author}{Yoneda, J.} \emph{et~al.}
\newblock \bibinfo{title}{A quantum-dot spin qubit with coherence limited by
  charge noise and fidelity higher than 99.9{\%}}.
\newblock \emph{\bibinfo{journal}{Nature Nanotechnology}}
  \textbf{\bibinfo{volume}{13}}, \bibinfo{pages}{102--106}
  (\bibinfo{year}{2017}).

\bibitem{Maurand2016}
\bibinfo{author}{Maurand, R.} \emph{et~al.}
\newblock \bibinfo{title}{A {CMOS} silicon spin qubit}.
\newblock \emph{\bibinfo{journal}{Nature Communications}}
  \textbf{\bibinfo{volume}{7}}, \bibinfo{pages}{13575} (\bibinfo{year}{2016}).

\bibitem{Crippa2018}
\bibinfo{author}{Crippa, A.} \emph{et~al.}
\newblock \bibinfo{title}{Electrical spin driving by g -matrix modulation in
  spin-orbit qubits}.
\newblock \emph{\bibinfo{journal}{Physical Review Letters}}
  \textbf{\bibinfo{volume}{120}}, \bibinfo{pages}{137702}
  (\bibinfo{year}{2018}).

\bibitem{Watzinger2018}
\bibinfo{author}{Watzinger, H.} \emph{et~al.}
\newblock \bibinfo{title}{A germanium hole spin qubit}.
\newblock \emph{\bibinfo{journal}{Nature Communications}}
  \textbf{\bibinfo{volume}{9}}, \bibinfo{pages}{3902} (\bibinfo{year}{2018}).

\bibitem{Hendrickx2020a}
\bibinfo{author}{Hendrickx, N.~W.} \emph{et~al.}
\newblock \bibinfo{title}{A single-hole spin qubit}.
\newblock \emph{\bibinfo{journal}{Nature Communications}}
  \textbf{\bibinfo{volume}{11}}, \bibinfo{pages}{3478} (\bibinfo{year}{2020}).

\bibitem{Hendrickx2020}
\bibinfo{author}{Hendrickx, N.~W.}, \bibinfo{author}{Franke, D.~P.},
  \bibinfo{author}{Sammak, A.}, \bibinfo{author}{Scappucci, G.} \&
  \bibinfo{author}{Veldhorst, M.}
\newblock \bibinfo{title}{Fast two-qubit logic with holes in germanium}.
\newblock \emph{\bibinfo{journal}{Nature}} \textbf{\bibinfo{volume}{577}},
  \bibinfo{pages}{487--491} (\bibinfo{year}{2020}).

\bibitem{Froning2021}
\bibinfo{author}{Froning, F. N.~M.} \emph{et~al.}
\newblock \bibinfo{title}{Ultrafast hole spin qubit with gate-tunable
  spin{\textendash}orbit switch functionality}.
\newblock \emph{\bibinfo{journal}{Nature Nanotechnology}}
  \textbf{\bibinfo{volume}{16}}, \bibinfo{pages}{308--312}
  (\bibinfo{year}{2021}).

\bibitem{Wang2020}
\bibinfo{author}{Wang, K.} \emph{et~al.}
\newblock \bibinfo{title}{Ultrafast operations of a hole spin qubit in ge
  quantum dot}.
\newblock \emph{\bibinfo{journal}{ArXiV}}  (\bibinfo{year}{2020}).
\newblock \eprint{2006.12340}.

\bibitem{Bosco2021}
\bibinfo{author}{Bosco, S.}, \bibinfo{author}{Benito, M.},
  \bibinfo{author}{Adelsberger, C.} \& \bibinfo{author}{Loss, D.}
\newblock \bibinfo{title}{Squeezed hole spin qubits in ge quantum dots with
  ultrafast gates at low power}.
\newblock \emph{\bibinfo{journal}{ArXiv}}  (\bibinfo{year}{2021}).
\newblock \eprint{2103.16724}.

\bibitem{Wang2021}
\bibinfo{author}{Wang, Z.} \emph{et~al.}
\newblock \bibinfo{title}{Optimal operation points for ultrafast, highly
  coherent ge hole spin-orbit qubits}.
\newblock \emph{\bibinfo{journal}{npj Quantum Information}}
  \textbf{\bibinfo{volume}{7}}, \bibinfo{pages}{54} (\bibinfo{year}{2021}).

\bibitem{Jirovec2021}
\bibinfo{author}{Jirovec, D.} \emph{et~al.}
\newblock \bibinfo{title}{A singlet-triplet hole spin qubit in planar
  $\mathrm{Ge}$}.
\newblock \emph{\bibinfo{journal}{Nature Materials}}  (\bibinfo{year}{2021}).

\bibitem{Dickel2015}
\bibinfo{author}{Dickel, C.}, \bibinfo{author}{Foletti, S.},
  \bibinfo{author}{Umansky, V.} \& \bibinfo{author}{Bluhm, H.}
\newblock \bibinfo{title}{Characterization of $\mathrm{S-T_+}$ transition
  dynamics via correlation measurements}.
\newblock \emph{\bibinfo{journal}{Physical Review B}}
  \textbf{\bibinfo{volume}{92}}, \bibinfo{pages}{125402}
  (\bibinfo{year}{2015}).

\bibitem{Nichol2015}
\bibinfo{author}{Nichol, J.~M.} \emph{et~al.}
\newblock \bibinfo{title}{Quenching of dynamic nuclear polarization by
  spin{\textendash}orbit coupling in {GaAs} quantum dots}.
\newblock \emph{\bibinfo{journal}{Nature Communications}}
  \textbf{\bibinfo{volume}{6}}, \bibinfo{pages}{7682} (\bibinfo{year}{2015}).

\bibitem{Maune2012}
\bibinfo{author}{Maune, B.~M.} \emph{et~al.}
\newblock \bibinfo{title}{Coherent singlet-triplet oscillations in a
  silicon-based double quantum dot}.
\newblock \emph{\bibinfo{journal}{Nature}} \textbf{\bibinfo{volume}{481}},
  \bibinfo{pages}{344--347} (\bibinfo{year}{2012}).

\bibitem{HarveyCollard2019}
\bibinfo{author}{Harvey-Collard, P.} \emph{et~al.}
\newblock \bibinfo{title}{Spin-orbit interactions for singlet-triplet qubits in
  silicon}.
\newblock \emph{\bibinfo{journal}{Physical Review Letters}}
  \textbf{\bibinfo{volume}{122}}, \bibinfo{pages}{217702}
  (\bibinfo{year}{2019}).

\bibitem{Tanttu2019}
\bibinfo{author}{Tanttu, T.} \emph{et~al.}
\newblock \bibinfo{title}{Controlling spin-orbit interactions in silicon
  quantum dots using magnetic field direction}.
\newblock \emph{\bibinfo{journal}{Physical Review X}}
  \textbf{\bibinfo{volume}{9}}, \bibinfo{pages}{021028} (\bibinfo{year}{2019}).

\bibitem{Petta2005}
\bibinfo{author}{Petta, J.~R.}
\newblock \bibinfo{title}{Coherent manipulation of coupled electron spins in
  semiconductor quantum dots}.
\newblock \emph{\bibinfo{journal}{Science}} \textbf{\bibinfo{volume}{309}},
  \bibinfo{pages}{2180--2184} (\bibinfo{year}{2005}).

\bibitem{Petta2010}
\bibinfo{author}{Petta, J.~R.}, \bibinfo{author}{Lu, H.} \&
  \bibinfo{author}{Gossard, A.~C.}
\newblock \bibinfo{title}{A coherent beam splitter for electronic spin states}.
\newblock \emph{\bibinfo{journal}{Science}} \textbf{\bibinfo{volume}{327}},
  \bibinfo{pages}{669--672} (\bibinfo{year}{2010}).

\bibitem{Shevchenko2010}
\bibinfo{author}{Shevchenko, S.}, \bibinfo{author}{Ashhab, S.} \&
  \bibinfo{author}{Nori, F.}
\newblock \bibinfo{title}{Landau{\textendash}zener{\textendash}stückelberg
  interferometry}.
\newblock \emph{\bibinfo{journal}{Physics Reports}}
  \textbf{\bibinfo{volume}{492}}, \bibinfo{pages}{1--30}
  (\bibinfo{year}{2010}).

\bibitem{Marie1999}
\bibinfo{author}{Marie, X.} \emph{et~al.}
\newblock \bibinfo{title}{Hole spin quantum beats in quantum-well structures}.
\newblock \emph{\bibinfo{journal}{Physical Review B}}
  \textbf{\bibinfo{volume}{60}}, \bibinfo{pages}{5811--5817}
  (\bibinfo{year}{1999}).

\bibitem{Winkler2003}
\bibinfo{author}{Winkler, R.}
\newblock \emph{\bibinfo{title}{Spin{\textemdash}Orbit Coupling Effects in
  Two-Dimensional Electron and Hole Systems}} (\bibinfo{publisher}{Springer
  Berlin Heidelberg}, \bibinfo{year}{2003}).

\bibitem{Trifonov2021}
\bibinfo{author}{Trifonov, A.~V.} \emph{et~al.}
\newblock \bibinfo{title}{Strong enhancement of heavy-hole landé factor $q$ in
  ingaas symmetric quantum dots revealed by coherent optical spectroscopy}.
\newblock \emph{\bibinfo{journal}{ArXiv}}  (\bibinfo{year}{2021}).
\newblock \eprint{2103.13653}.

\bibitem{Stepanenko2012}
\bibinfo{author}{Stepanenko, D.}, \bibinfo{author}{Rudner, M.},
  \bibinfo{author}{Halperin, B.~I.} \& \bibinfo{author}{Loss, D.}
\newblock \bibinfo{title}{Singlet-triplet splitting in double quantum dots due
  to spin-orbit and hyperfine interactions}.
\newblock \emph{\bibinfo{journal}{Physical Review B}}
  \textbf{\bibinfo{volume}{85}}, \bibinfo{pages}{075416}
  (\bibinfo{year}{2012}).

\bibitem{Fischer2008}
\bibinfo{author}{Fischer, J.}, \bibinfo{author}{Coish, W.~A.},
  \bibinfo{author}{Bulaev, D.~V.} \& \bibinfo{author}{Loss, D.}
\newblock \bibinfo{title}{Spin decoherence of a heavy hole coupled to nuclear
  spins in a quantum dot}.
\newblock \emph{\bibinfo{journal}{Physical Review B}}
  \textbf{\bibinfo{volume}{78}}, \bibinfo{pages}{155329}
  (\bibinfo{year}{2008}).

\bibitem{Mutter2021}
\bibinfo{author}{Mutter, P.~M.} \& \bibinfo{author}{Burkard, G.}
\newblock \bibinfo{title}{Pauli spin blockade with site-dependent g tensors and
  spin-polarized leads}.
\newblock \emph{\bibinfo{journal}{Physical Review B}}
  \textbf{\bibinfo{volume}{103}}, \bibinfo{pages}{245412}
  (\bibinfo{year}{2021}).

\bibitem{Mutter2021ST}
\bibinfo{author}{Mutter, P.~M.} \& \bibinfo{author}{Burkard, G.}
\newblock \bibinfo{title}{All-electrical control of hole singlet-triplet spin
  qubits at low leakage points}.
\newblock \emph{\bibinfo{journal}{ArXiv}}  (\bibinfo{year}{2021}).
\newblock \eprint{2107.12622}.

\bibitem{Barthel2012}
\bibinfo{author}{Barthel, C.} \emph{et~al.}
\newblock \bibinfo{title}{Relaxation and readout visibility of a
  singlet-triplet qubit in an overhauser field gradient}.
\newblock \emph{\bibinfo{journal}{Physical Review B}}
  \textbf{\bibinfo{volume}{85}} (\bibinfo{year}{2012}).

\bibitem{Granger2015}
\bibinfo{author}{Granger, G.} \emph{et~al.}
\newblock \bibinfo{title}{Visibility study of $\mathrm{S-T_+}$
  landau-zener-stückelberg oscillations without applied initialization}.
\newblock \emph{\bibinfo{journal}{Physical Review B}}
  \textbf{\bibinfo{volume}{91}}, \bibinfo{pages}{115309}
  (\bibinfo{year}{2015}).

\bibitem{Orona2018}
\bibinfo{author}{Orona, L.~A.} \emph{et~al.}
\newblock \bibinfo{title}{Readout of singlet-triplet qubits at large magnetic
  field gradients}.
\newblock \emph{\bibinfo{journal}{Physical Review B}}
  \textbf{\bibinfo{volume}{98}}, \bibinfo{pages}{125404}
  (\bibinfo{year}{2018}).

\bibitem{Danon2009}
\bibinfo{author}{Danon, J.} \& \bibinfo{author}{Nazarov, Y.~V.}
\newblock \bibinfo{title}{Pauli spin blockade in the presence of strong
  spin-orbit coupling}.
\newblock \emph{\bibinfo{journal}{Physical Review B}}
  \textbf{\bibinfo{volume}{80}}, \bibinfo{pages}{041301(R)}
  (\bibinfo{year}{2009}).

\bibitem{Golovach2004}
\bibinfo{author}{Golovach, V.~N.}, \bibinfo{author}{Khaetskii, A.} \&
  \bibinfo{author}{Loss, D.}
\newblock \bibinfo{title}{Phonon-induced decay of the electron spin in quantum
  dots}.
\newblock \emph{\bibinfo{journal}{Physical Review Letters}}
  \textbf{\bibinfo{volume}{93}}, \bibinfo{pages}{016601}
  (\bibinfo{year}{2004}).

\end{thebibliography}
\pagebreak
\widetext
\newpage
\begin{center}
\textbf{\large Supplementary: Hole spin orbit physics in quantum dot systems with large g-factor differences}
\end{center}

\setcounter{equation}{0}
\setcounter{figure}{0}
\setcounter{table}{0}
\setcounter{page}{1}
\makeatletter
\renewcommand{\theequation}{S\arabic{equation}}
\renewcommand{\thefigure}{S\arabic{figure}}
\renewcommand{\bibnumfmt}[1]{[S#1]}
\renewcommand{\citenumfont}[1]{S#1}

\section{Stability diagram, pulsing, tunnel coupling}
\begin{figure}
    \centering
    \includegraphics[width = 0.5\textwidth]{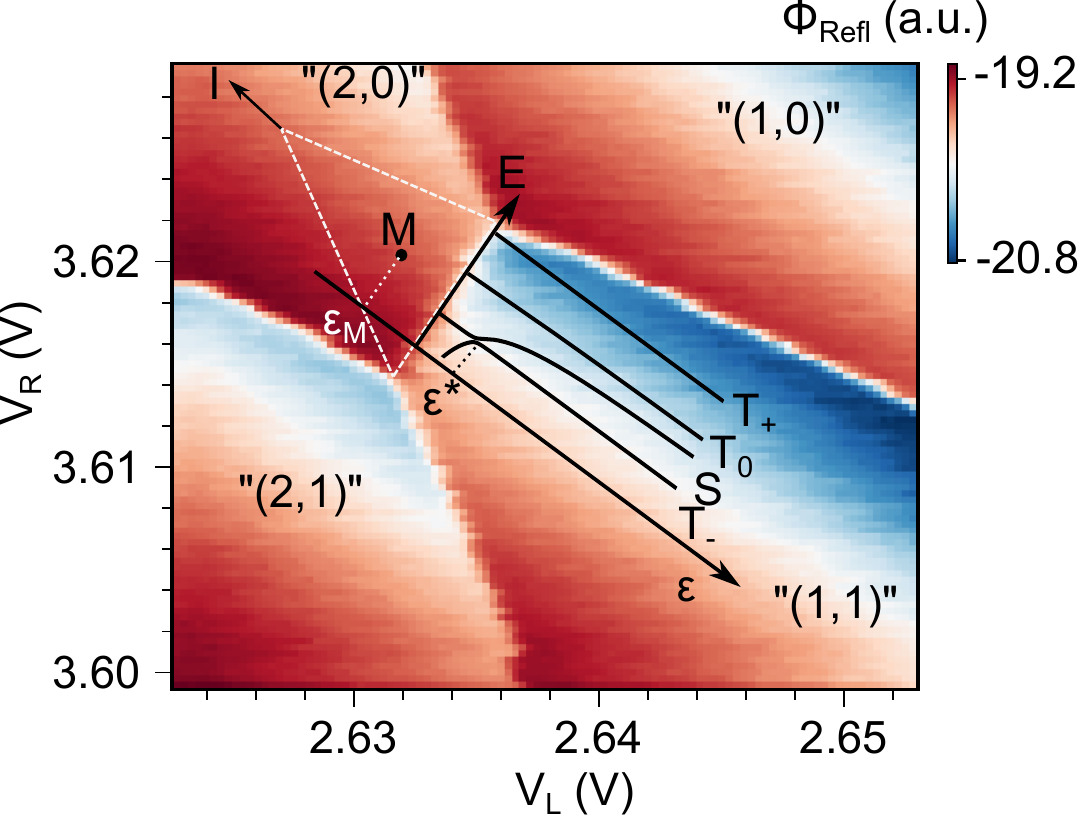}
    \caption{ The effective hole numbers are labelled ``($n_L$, $n_R$)''. The real hole number is ($n_L+2$, $2n+n_R$) since we do not know the exact number of holes in the right dot. The dashed triangle marks the region in which Pauli spin blockade occurs and the measurement point (M) is located at the center of it at $\epsilon = \epsilon_M$. ``I'' marks the initialization point. The dispersion relation is overlaid on top of the ``(1,1)'' region. The avoided $S-T_-$ crossing occurs at $\epsilon = \epsilon^*$}
    \label{fig:StabilityDiagram}
\end{figure}

The stability diagram around the transition of interest is reported in Fig. \ref{fig:StabilityDiagram}. The effective hole numbers are labelled ``($n_L$, $n_R$)'' while the real hole number is ($n_L+2$, $n_R+2n$). The dispersion relation is overlaid and highlights the 4 states involved in the dynamics of the system. When pulsing we initialize (I) deep in ``(2,0)'' with a short pulse knowing that the short relaxation time $T_1$ ensures decay to a singlet. Only for the in-plane field direction this is not true as is further discussed below. 

\begin{figure}
    \centering
    \includegraphics[width = \textwidth]{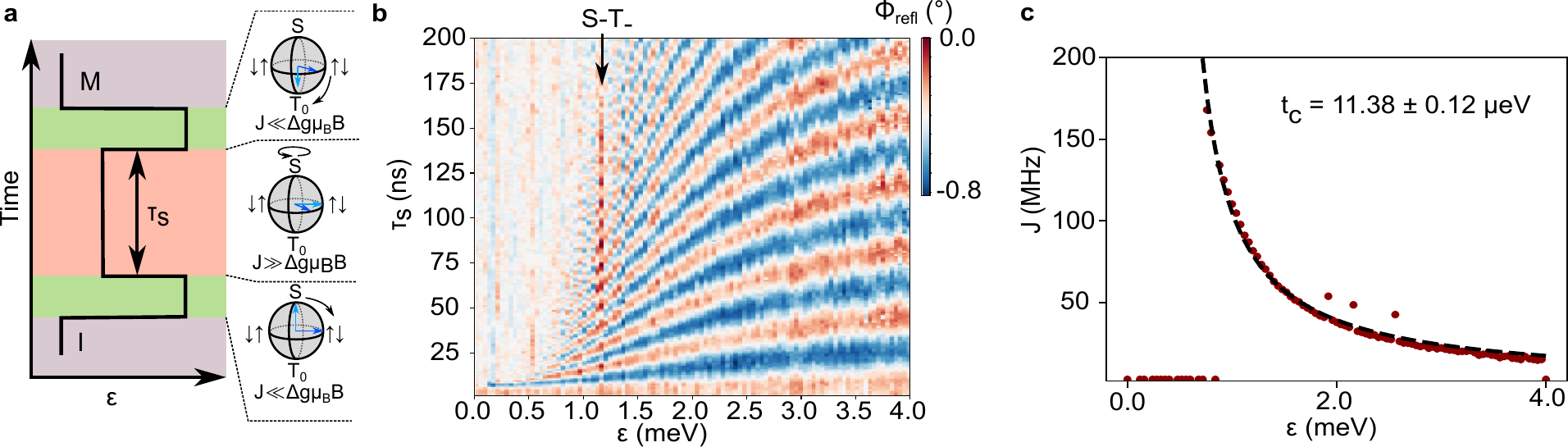}
    \caption{a) The pulse sequence to obtain exchange oscillations features two short pulses at high $\epsilon$ (green) where the exchange interaction $J \ll \Delta g \mu_B B$. Here, the system rotates around the x-axis of the Bloch sphere. The exchange interaction can be turned on by pulsing to lower $\epsilon$ (orange). Here $J \gg g_\perp^- \mu_B B$ and the system evolves around the z-axis of the Bloch sphere. b) The exchange oscillations as a function of $\epsilon$ and separation time $\tau_S$ at $B_\perp = \SI{1}{\milli\tesla}$ display the expected frequency increase for lower detuning as the exchange interaction becomes dominant. Around $\SI{1.2}{\milli e \volt}$ the $S-T_-$ degeneracy is visible as an increase in triplet return signal (red). 
    c) The magnitude of $J$ in $\SI{}{\mega\hertz}$ can be fit by $J(\epsilon, t_C) = \sqrt{\frac{\epsilon^2}{4}+2t_C^2}-\frac{\epsilon}{2}$ (black dashed line) and we extract $t_C = \SI{11.38}{\micro e \volt}$.}
    \label{fig:Exchange}
\end{figure}

We extract the tunnel coupling from exchange oscillations. A typical pulse sequence to probe the exchange interaction $J$ is depicted in the inset of Fig. \ref{fig:Exchange}a. After initialization in $S(2,0)$ the system is pulsed rapidly to large $\epsilon$ and left to evolve for a time corresponding to a $\pi/2$ rotation around the x-axis of the Bloch-sphere. A subsequent pulse at lower $\epsilon$ for a time $\tau_S$ increases the exchange interaction $J(\epsilon) = \sqrt{\frac{\epsilon^2}{4}+2t_C^2}-\frac{\epsilon}{2}$. Therefore, the oscillation frequency $f = \sqrt{J(\epsilon)^2+(g_\perp^-\mu_BB)^2}$ will reflect the change in $J$ and increase for lower $\epsilon$. The fit in Fig. \ref{fig:Exchange}, therefore, allows to extract the tunnel coupling. As the coherence time for very low $\epsilon$ tends to 0, all the points at $\epsilon<\epsilon^*$ are not considered in the fit. $\epsilon^*$ is highlighted by the black arrow where the $S-T_-$ mixing is prominently visible.

\section{\label{sec:SmallAngles}Singlet-triplet dynamics for small angles }

\begin{figure}
    \centering
    \includegraphics[width = 0.9\textwidth]{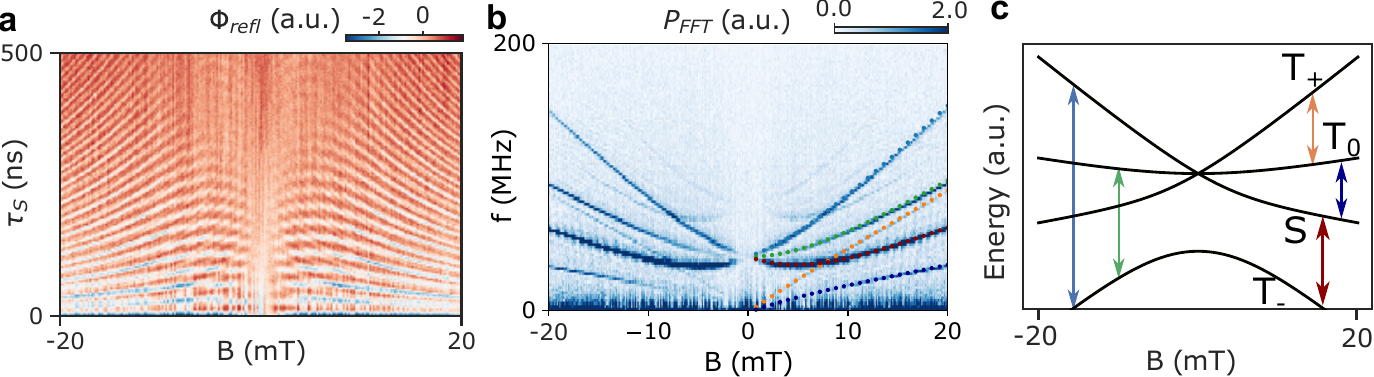}
    \caption{a)  Singlet oscillations as a function of in-plane magnetic field and free evolution time under fast pulsing revealing a complicated oscillation pattern.  b) A FFT of a) unveils many frequency components. The dotted lines are fits to the transition frequencies highlighted in c). We find good agreement between the data and the model. An additional set of lines appears at slightly higher frequencies which we attribute to spectral leakage due to the imperfect initialization \cite{Granger2015}. c) The arrows point out the transition frequencies between the eigenstates for the in-plane dispersion relation.}
    \label{fig:RabiInPlane}
\end{figure}

\begin{figure}
    \centering
    \includegraphics{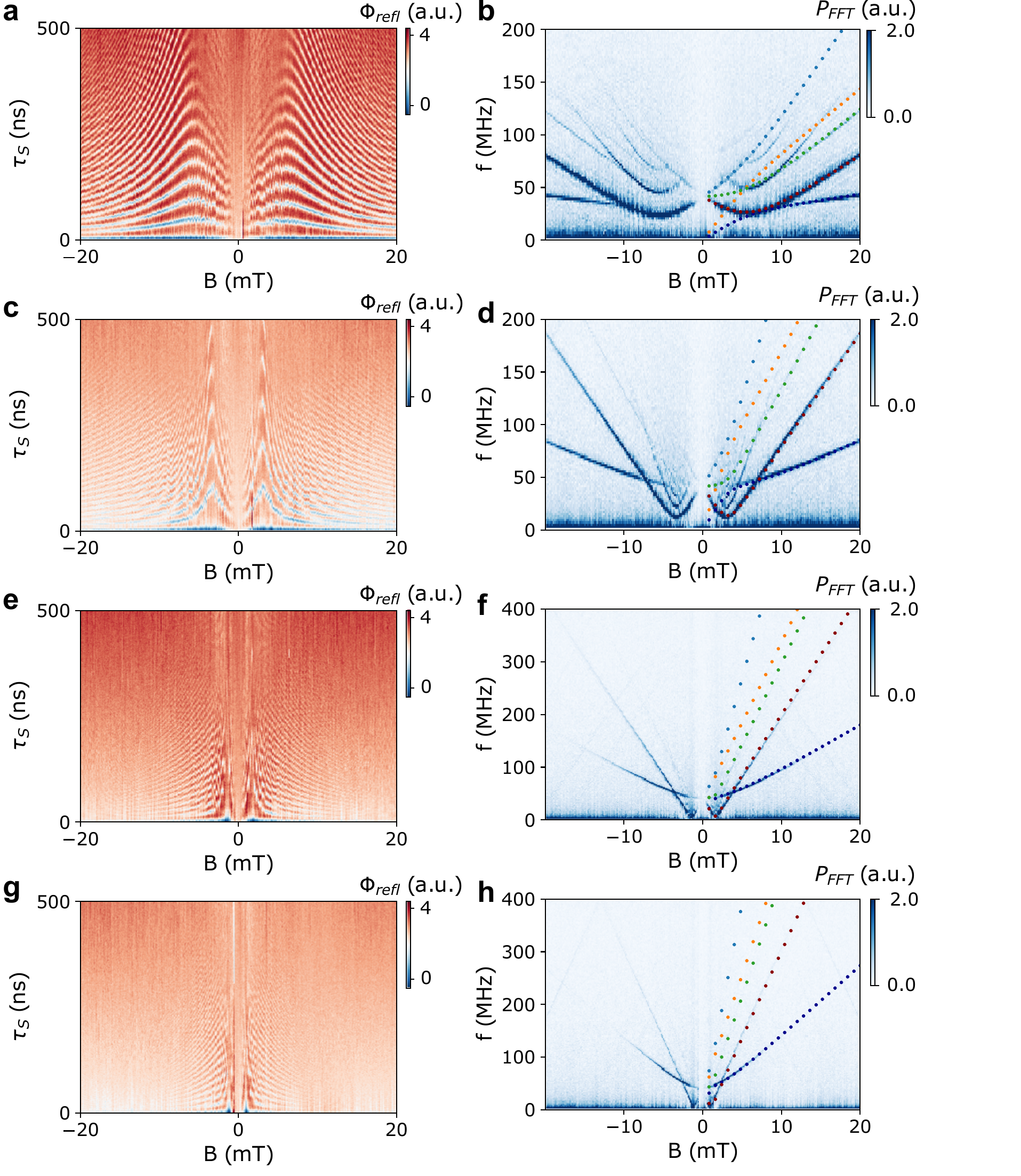}
    \caption{Singlet-triplet oscillations as a function of B at different $\theta$. a) Oscillations at $\theta = \SI{5}{\degree}$. b) FFT of a) with overlaid frequency lines for positive fields as extracted from the model. c) Oscillations at $\theta = \SI{10}{\degree}$. d) FFT of c) with overlaid frequency lines as extracted from the model. e) Oscillations at $\theta = \SI{20}{\degree}$. f) FFT of e) with overlaid frequency lines as extracted from the model. g) Oscillations at $\theta = \SI{30}{\degree}$. h) FFT of g) with overlaid frequency lines as extracted from the model.}
    \label{fig:RabivsThetaSupp}
\end{figure}

We probe singlet-triplet oscillations at small angles by rapidly pulsing to large $\epsilon$ and observing the signal of the returned state. In Fig. \ref{fig:RabiInPlane}a the resulting oscillation pattern is displayed for $\theta = \SI{0}{\degree}$. The FFT in \ref{fig:RabiInPlane}b reveals several frequency components. The dotted lines are extracted from our model and the corresponding dispersion relation is depicted in \ref{fig:RabiInPlane}c. The colored arrows mark the transitions we actually observe in the FFT plot. The singlet-triplet oscillation and FFT plots for $\theta = 5,10,20$ and $\SI{30}{\degree}$ are further depicted in Fig. \ref{fig:RabivsThetaSupp}. We generally find a good agreement with our theoretical model. 

\newpage

\section{$S-T_-$ degeneracy as a function of magnetic field direction}
\begin{figure}
    \centering
    \includegraphics[width = \textwidth]{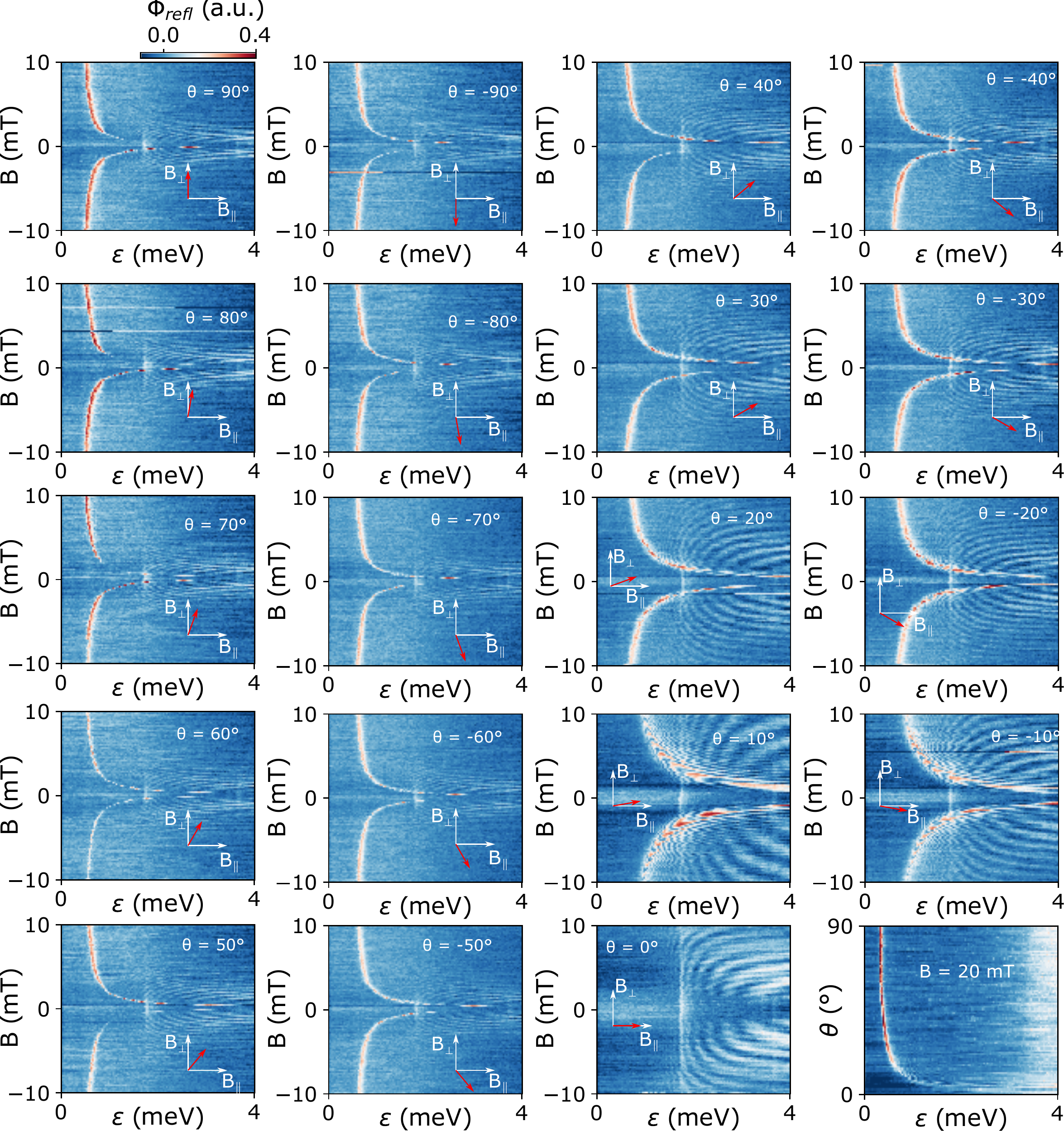}
    \caption{$S-T_-$ degeneracy funnel for different angles for $\tau_S = \SI{250}{\nano\second}$. Clearly the funnel is well defined for out-of-plane fields and evolves into an oscillation pattern for in-plane fields. This suggests a dependence stemming from the g-factor anisotropy of HH states. In fact, our model fits the experimentally observed funnel shapes with $g_\perp^+ = 12$, $g_\perp^- = 2.05$, $g_\parallel^+ = 0.1$, $g_\parallel^- = 0.43$. Moreover, the spin-orbit vector pointing along $\theta = \SI{60}{\degree}$ gives rise to an asymmetry between the sharpness of the funnel at $+\SI{60}{\degree}$ and $-\SI{60}{\degree}$, also confirmed by single Landau-Zener sweeps. The bottom right plot shows the $S-T_-$ degeneracy at $\SI{20}{\milli\tesla}$ for positive angles. This is used to calibrate the pulse amplitude necessary to overcome the $S-T_-$ degeneracy in the single Landau-Zener sweeps.}
    \label{fig:SuppFunnelvsAngle}
    \end{figure}

We report the $S-T_-$ degeneracy for different magnetic field directions in Fig. \ref{fig:SuppFunnelvsAngle}. For the out-of-plane direction the typical funnel shape can clearly be observed. Towards the in-plane direction the $S-T_-$ degeneracy evolves into an oscillation pattern suggesting a large coupling term between $S$ and $T_-$. This anisotropy, which follows the g-factor anisotropy of confined HH states, indicates a dependence related to the Zeeman terms in $H_{tot}$. Indeed, our model can reproduce the plots assuming $g_\parallel^- > g_\parallel^+$. 

\begin{figure}
    \centering
    \includegraphics{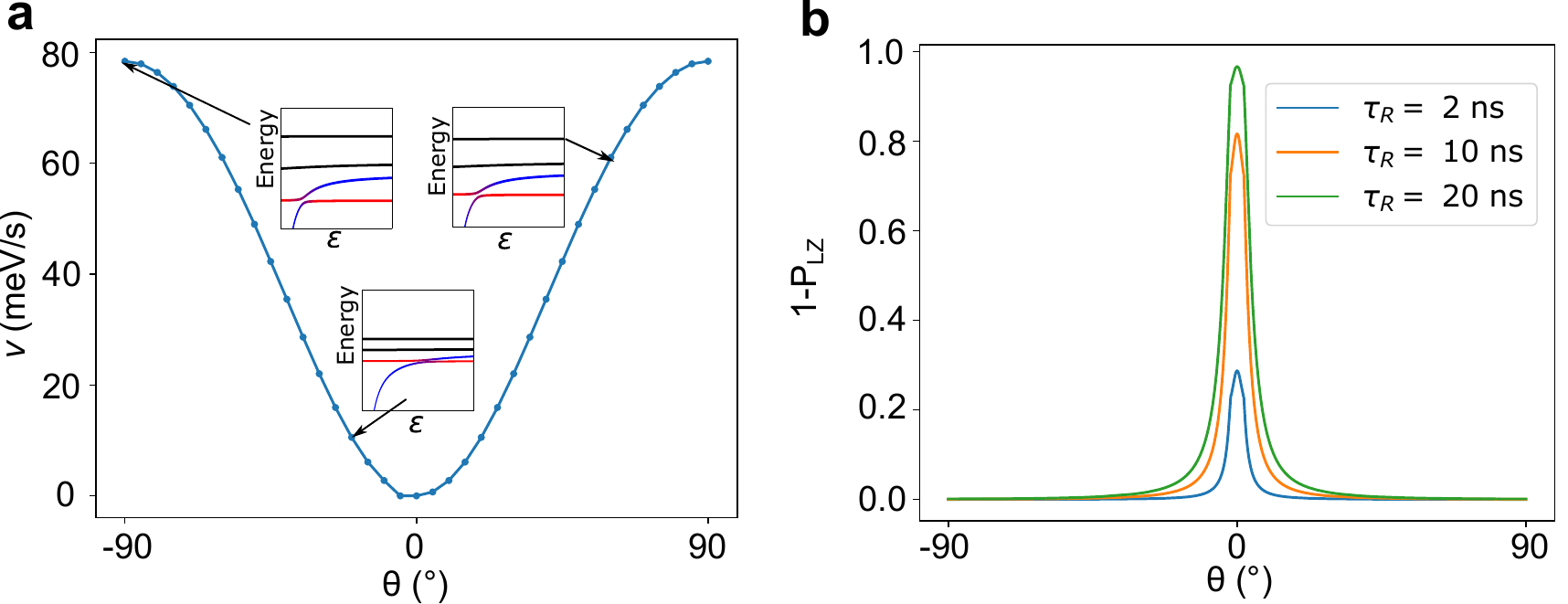}
    \caption{a) Velocity $v$ as a function of $\theta$ for a magnetic field strength of $\SI{20}{\milli\tesla}$. Because of the g-factor anisotropy the Zeeman energy in the in-plane direction is lower. Consequently the velocity is reduced as $|dJ/d\epsilon|_{\epsilon = \epsilon^*}$ becomes smaller. The inset shows the simulated energy dispersion for the angles highlighted by the black arrows. For in-plane fields the avoided crossing clearly shifts to larger $\epsilon$ thereby reducing the effective level velocity.  b) $1-P_{LZ}$ as a function of $\theta$ for different ramp times. Hence, the g-factor difference in the in-plane direction allows fast and high fidelity initialization for single spin qubits. }
    \label{fig:VvsTheta}
\end{figure}

We furthermore report the effective level velocity $v = |\frac{dJ(\epsilon)}{d\epsilon}|_{\epsilon=\epsilon^*}\frac{\Delta\epsilon}{\tau_R}$ with $\Delta\epsilon = \SI{4}{\milli e \volt}$, $\tau_R = \SI{2}{\nano\second}$ as a function of the magnetic field angle $\theta$ for a magnetic field strength of $\SI{20}{\milli\tesla}$ (Fig. \ref{fig:VvsTheta}). Towards the in-plane direction the velocity is reduced as the Zeeman splitting is lowered due to the g-factor anisotropy. As a result, for small angles, the shortest possible ramp time does not allow a diabatic passage over the avoided crossing as pointed out in Fig. \ref{fig:TheoryFig}b. Incidentally, this is a favorable property for single spin qubits where fast initialization in the $\ket{\downarrow\downarrow}$ state is required. Indeed, for $\theta = 0$ the ground state in (1,1) can be initialized with fidelity $F = 1-P_{LZ}\approx 1 $ at a ramp time of 20 ns (Fig. \ref{fig:VvsTheta}). 

\section{Decay of $T_0$ state}
\begin{figure}
    \centering
    \includegraphics[width = 0.5\textwidth]{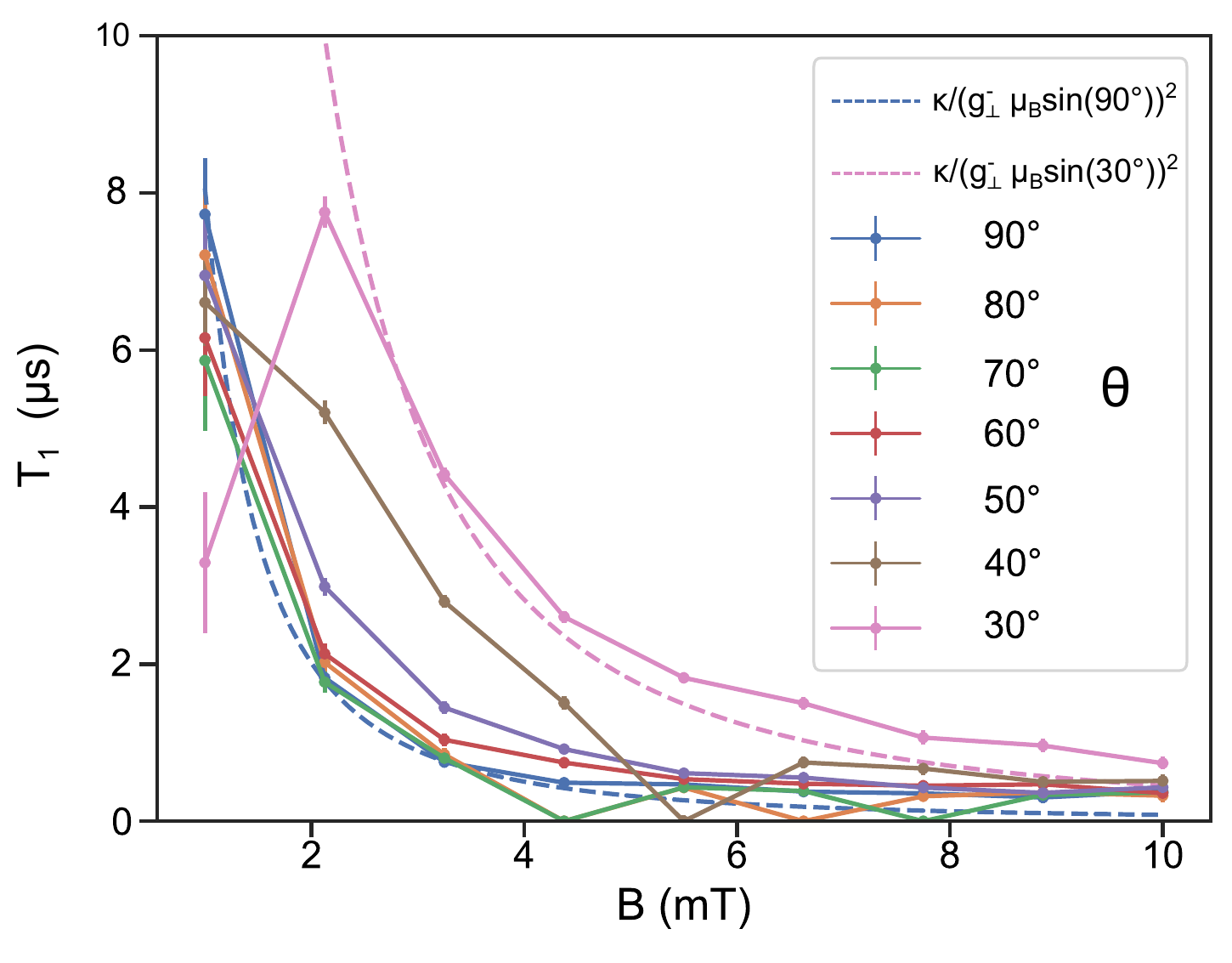}
    \caption{Relaxation time of the $\ket{T_0}$ state as a function of magnetic field strength and angle. In contrast to the decay of the $T_-$ state the relaxation of the $T_0$ state is proportional to $\kappa/\Delta E_Z^2$ (dashed lines) wiht $\kappa$ a proportionality constant. $T_1$ generally decreases for larger magnetic fields. Only for $\theta = \SI{30}{\degree}$ the relaxation time maximizes at $\SI{2}{\milli\tesla}$. This occurs because for lower magnetic fields the presence of the $S-T_-$ avoided crossing lowers the $T_0$ return probability as a diabatic passage can not be achieved with the finite rise time of our AWG. For angles $\theta < \SI{30}{\degree}$ no diabatic passage can be achieved. Therefore we cannot extract $T_1$ of the $T_0$ state.  }
    \label{fig:T1ofT0}
\end{figure}

The relaxation time of the $T_0$ state does not depend on the SOI (Eq. \ref{eq:effective_SO_field})and, therefore, displays a different anisotropy than the relaxation for the $T_-$ state. To probe this, we perform diabatic pulses and fit the decay of the return signal to $\Phi_{refl}(\tau_M = 0)\exp(-\frac{\tau_M}{T_1})$. As shown in Fig. \ref{fig:T1ofT0} the relaxation time follows $\approx \frac{1}{g_\perp^-\mu_B B\sin\theta}^2$ and is generally $< \SI{10}{\micro\second}$ making read-out at large fields challenging. Therefore, shelved read-out \cite{Orona2018} could be harvested where the read-out distinguishes between $S$ and $T_-$. At $\theta = \SI{60}{\degree}$ the relaxation of the $T_-$ state is maximised offering a sweet spot for shelved read-out.   

\section{Initialization and spectral leakage}
\begin{figure}
    \centering
    \includegraphics[width = \textwidth]{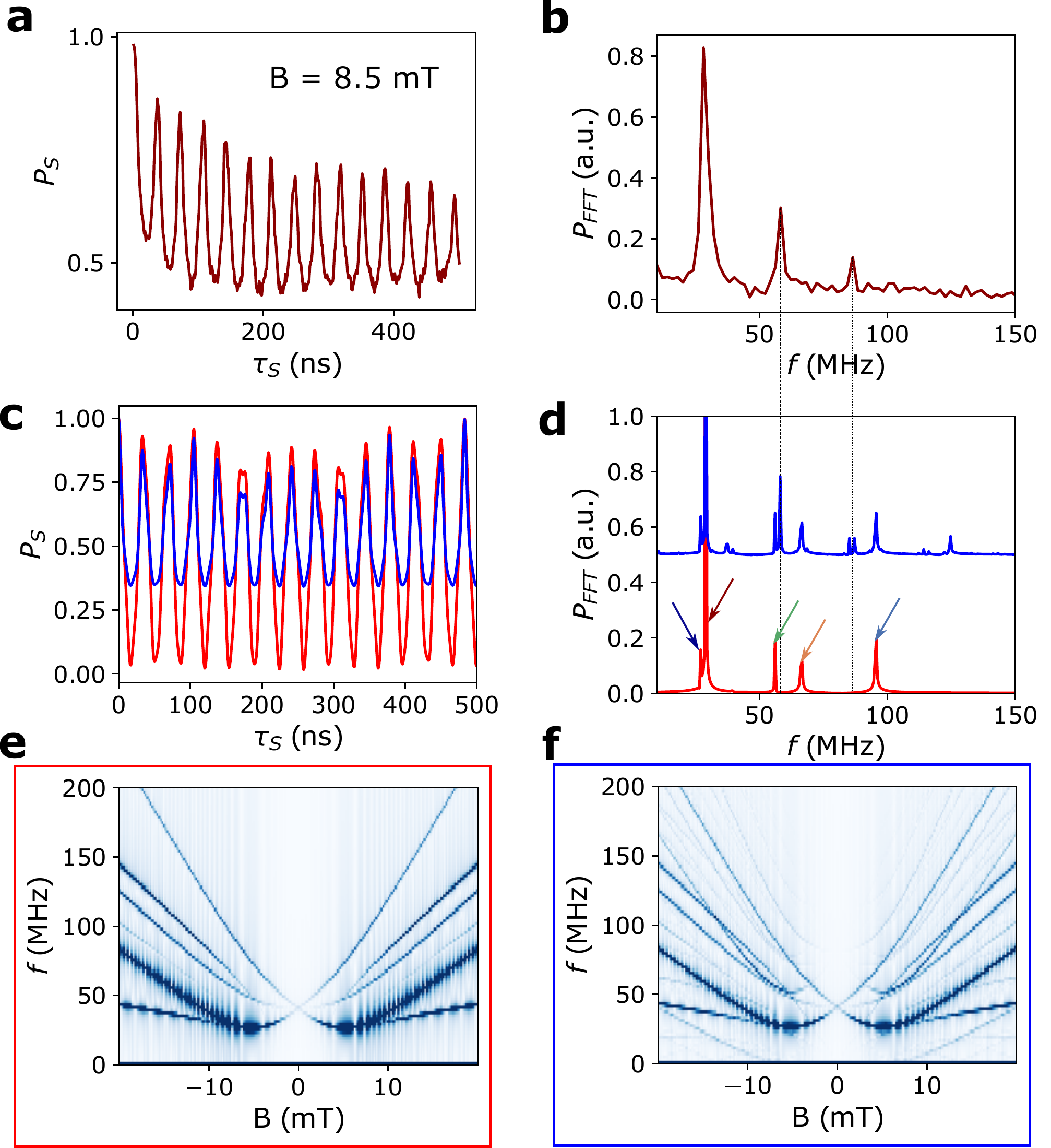}
    \caption{a) Line trace at 8 mT from Fig. \ref{fig:RabivsThetaSupp}a featuring a non-sinusoidal oscillation pattern. b) FFT of a) highlighting its frequency components. c) Simulated time evolution for the conditions in a) with proper initialization in a singlet (red) and improper initialization (blue). d) FFT of c) highlighting the frequency components of the blue and red trace (blue is offset for clarity). Comparing the 2 curves we notice additional frequency components arising in the blue one which can be mapped to the observed spectral peaks of b). The red and blue trace, however, both display peaks at the transition frequencies expected from our model. e) FFT of the simulated time evolution with proper initialization. All the transitions highlighted in Fig. \ref{fig:RabiInPlane}c are visible. f) FFT of the simulated time evolution under improper initialization. The additional frequency lines resemble what we measure in Fig. \ref{fig:RabivsThetaSupp}b.}
    \label{fig:init}
\end{figure}

The fast Fourier transform plots in Fig. \ref{fig:RabiInPlane} and Fig. \ref{fig:RabivsThetaSupp} display additional frequency components not predicted by our model which are especially evident at $\theta = \SI{5}{\degree}$ and $\SI{10}{\degree}$ and can be attributed to improper initialization of the singlet state prior to a pulse. In \cite{Granger2015} a similar system is analyzed and a general formula for the singlet return probability $P_S$ of non-properly initialized state (a superposition of $S$ and $T_-$) is extracted as
\begin{equation}
    P_S = P_S'P+[1-P_S'](1-P),
    \label{eq: Ps}
\end{equation}
where $P$ is the return probability with correct singlet initialization and 
\begin{equation}
    P_S' = \frac{P\exp(-\frac{\tau_M}{T_1})-1}{(2P-1)\exp(-\frac{\tau_M}{T_1})-1}.
    \label{eq:psprime}
\end{equation}
Here, $\tau_M$ is the measurement time and $T_1$ is the relaxation time of the $T_-$ state in the read-out point. It is clear that for $\tau_M \gg T_1$ we find $P_S = P$. However, when $\tau_M \approx T_1$ the singlet return probability deviates significantly from $P$. In Fig. \ref{fig:init}a we show a line trace of Fig. \ref{fig:RabivsThetaSupp} at $\theta = \SI{5}{\degree}$ and $B = \SI{8.5}{\milli\tesla}$. We clearly see that the oscillations are not sinusoidal. Fig. \ref{fig:init}b displays the FFT of the line trace and 3 distinct peaks can be observed. However, more peaks are present but buried in the background. Fig. \ref{fig:init}c shows a simulation of the time evolution of the state. The red trace assumes perfect initialization whereas the blue trace depicts the corrected singlet return probability according to Eq. \ref{eq: Ps} and \ref{eq:psprime} with
$\tau_M \approx T_1$. The blue trace indeed resembles what we measure. Furthermore, we inspect the FFT of the simulated data and find that indeed for the improperly initialized curve we observe additional frequency components. In the red trace we find only the frequency components corresponding to the transitions highlighted in Fig. \ref{fig:RabiInPlane} of the main text. The FFT plot extracted from the simulation of the properly initialized state is depicted as a function of magnetic field in Fig. \ref{fig:init}e. On the other hand, the improperly initialized simulated FFT is plotted in Fig. \ref{fig:init}f. Here, several additional lines are clearly visible and follow what we measure in Fig. \ref{fig:RabivsThetaSupp}b albeit with altered Fourier amplitudes.

\section{The model}
\label{sec:model}
We consider a tunnel coupled double quantum dot in the presence of an external magnetic field and the spin-orbit interaction (SOI). The system is modelled by the Hamiltonian
	\begin{align}
	\label{eq:total_Hamiltonian}
		H_{tot} = H_{orb} + H_Z + H_{SO},
	\end{align}
where $H_Z$ is the Zeeman Hamiltonian and $H_{SO}$ describes the SOI. The orbital part $H_{orb}$ reads
	\begin{align}
	\label{eq:orbital_Hamiltonian}
		H_{orb} = \epsilon \vert S_{20} \rangle \langle S_{20} \vert + \sqrt{2} t_C \left( \vert S_{11} \rangle \langle S_{20} \vert  + \vert S_{20} \rangle \langle S_{11} \vert   \right),
	\end{align}
where $\vert S_{20} \rangle$ is the singlet with double occupancy in the left dot which is detuned by an amount $\epsilon$ from the $(1,1)$ configuration, $t_C$ is the tunnel element in the one-particle picture, and $\vert S_{11} \rangle$ is the singlet with one hole in each dot. Diagonalizing~\eqref{eq:orbital_Hamiltonian} yields the hybridized singlet states
	\begin{subequations}
	\begin{align}
		\vert S_E \rangle = \cos \left( \frac{\Omega }{2} \right) \vert S_{20} \rangle   + \sin \left( \frac{\Omega }{2} \right)  \vert S_{11} \rangle, \\
		\vert S_G \rangle = \cos \left( \frac{\Omega }{2} \right) \vert S_{11} \rangle   - \sin \left( \frac{\Omega }{2} \right) \vert S_{20} \rangle,
	\end{align}
	\end{subequations}
where $\Omega = \arctan (2 \sqrt{2}t_C/\epsilon)$ is the orbital mixing angle and the corresponding energies are $E(S_{E/G}) = \epsilon/2 \pm \sqrt{\epsilon^2/4 + 2 t_C^2}$. In the following we consider a regime where the excited singlet is far detuned and may be neglected. Defining $\vert S \rangle \equiv \vert S_G \rangle$, we may focus on the four dimensional space spanned by the states $\vert S \rangle$,  $\vert T_+ \rangle$, $\vert T_0 \rangle$ and $\vert T_- \rangle$, where $\vert T_{0, \pm} \rangle$ denote the triplet states in the (1,1) configuration.

For a magnetic field of magnitude $B$ and angle $\theta$ as measured from the plane defined by the heterostructure, the Zeeman Hamiltonian $H_Z$ reads,
	\begin{align}
	\label{eq:Zeeman_Hamiltonian}
	\begin{split}
		H_Z =   \frac{\mu_B B }{2} \Bigg\lbrace \sin \theta \left[ \sum_{\pm} \pm g_{\perp}^+  \vert T_{\pm} \rangle \langle T_{\pm} \vert  +  g_{\perp}^- \left( \vert S \rangle \langle T_0 \vert + \vert T_0 \rangle \langle S \vert \right) \cos \left( \frac{\Omega }{2} \right) \right] \\
		+  \frac{\cos \theta}{ \sqrt{2}} \left[  \sum_{\pm} \left( g_{\parallel}^+  \vert T_0 \rangle \langle T_{\pm} \vert \mp   g_{\parallel}^-  \vert S \rangle \langle T_{\pm} \vert \cos \left( \frac{\Omega }{2} \right) + \text{H.c.} \right) \right] \Bigg\rbrace,
	\end{split}
	\end{align}
where $g_j^{\pm} = g_j^L \pm g_j^R$. Note that the axes are labelled such that $y$ is the double quantum dot axis, and the magnetic field is applied in the $x$-$z$-plane.

Finally, we consider a SOI parametrized by an in-plane spin-orbit vector, $\mathbf{t}_{SO} = (t_x, t_y, 0)$, where all entries are real. Such in-plane spin-flip tunneling terms stem from the cubic Rashba SOI~\cite{Mutter2021}, while this type of SOI does not induce out-of-plane terms $t_z$. Consequently, the spin-orbit Hamiltonian $H_{SO}$ reads,
	\begin{align}
	\label{eq:H_SO_spin_orbit_vector}
		H_{SO} =  \sin \left( \frac{\Omega }{2} \right) \sum_{\pm} (t_x \pm i t_y) \vert T_{\pm} \rangle \langle S \vert + \text{H.c.},
	\end{align}
An appropriate basis for studying the singlet-triplet anticrossing is given by the eigenstates of the total Hamiltonian~\eqref{eq:total_Hamiltonian} for equal g-factors and in the absence of the SOI. Transforming the total Hamiltonian in the presence of the SOI and g-factor differences into this basis, we find for the singlet-triplet splitting,
	\begin{align}
	\label{eq:Delta_ST-}
		\Delta_{ST_-} = \left\vert \Delta_{SO} \sin \left( \frac{\Omega }{2} \right)  + \Delta E_Z \cos \left( \frac{\Omega }{2} \right)  \right\vert,
	\end{align}
where the spin-orbit splitting $\Delta_{SO} $ and the Zeeman splitting $\Delta E_Z$ are given by
	\begin{subequations}
	\begin{align}
		& \Delta_{SO} = t_y - i t_x \frac{g_{\perp}^+ \sin \theta}{\sqrt{( g_{\parallel}^+ \cos \theta)^2 + (g_{\perp}^+ \sin \theta)^2}}, \\
		& \Delta E_Z = \frac{\mu_B B}{4 \sqrt{2}} \frac{\left( g_{\parallel}^- g_{\perp}^+ - g_{\parallel}^+ g_{\perp}^- \right) \sin (2 \theta)}{\sqrt{( g_{\parallel}^+ \cos \theta)^2 + (g_{\perp}^+ \sin \theta)^2}}.
	\end{align}
	\label{eq:DeltaEz}
	\end{subequations}
In Landau-Zener measurements the mixing angle $\Omega$ is a function of the detuning at the anticrossing between the singlet and the low-energy polarized triplet, which is described by the equation
	\begin{align}
	 	\frac{\epsilon}{2} - \sqrt{\frac{\epsilon^2}{4} + 2 t_C^2} = - \frac{\mu_B B}{2} G(\theta), 
	 \end{align}
where $G(\theta) = \sqrt{(g_{\parallel}^+ \cos \theta)^2 + (g_{\perp}^+ \sin \theta)^2}$. Rearranging gives the detuning at which the anticrossing is located,
	\begin{align}
		 \epsilon_* = \frac{8 t_C^2 - \left( \mu_B B G(\theta) \right)^2 }{2\mu_B B G(\theta) },
	\end{align}
and consequently the mixing angle at the anticrossing reads
	\begin{align}
		\Omega_* = \Omega(\epsilon_*) = \arctan \left( \frac{4 \sqrt{2} t_C \mu_B B G(\theta)}{8t_c^2 - (\mu_B B G(\theta))^2}\right).
	\end{align}
In Fig.~\ref{fig:TheoryFig}d we show a comparison between the analytical result for $\Delta_{ST_-}$ evaluated at the detuning at the anticrossing and simulated results obtained by exact numerical diagonalization of the total system Hamiltonian~\eqref{eq:total_Hamiltonian}. For the latter, we obtain the splitting as the minimum difference between the two lowest eigenenergies. We find excellent agreement for all angles except in a narrow region around $\theta = 0$. The dip predicted by the analytical formula is not present in the simulations. We attribute this deviation to the small in-plane Zeeman energies $\sim g_{\parallel}^+$ which violate the assumption of a well separated two-level system. In this case the splitting is not given by the simple expression~\eqref{eq:Delta_ST-} which describes the coupling between the ground state singlet and the low-energy polarized triplet. Instead, the contributions of the remaining two triplet states must be taken into account near $\theta=0$, and the coupling is altered. 

\section{Spin-orbit vector, field and length}
\label{sec:spin-orbit_vector_field_length}
In Eq.~\eqref{eq:H_SO_spin_orbit_vector} we model the SOI using the spin-orbit vector $\mathbf{t}_{\text{SO}} = (t_x, t_y, t_z)$~\cite{Danon2009} which contains genuine spin-flip tunneling terms. In our structure we consider an in-plane spin-orbit vector, but different forms with $t_z \neq 0$ are possible in other systems. The most general effective form in the basis $\mathcal{B}$ defined in Sec.~\ref{sec:model} is
	\begin{align}
	\label{eq:general_SOI}
		H_{SO} = \sin \left( \frac{\Omega }{2} \right) \left(i \sqrt{2} t_z \vert T_0 \rangle \langle S \vert + \sum_{\pm} (t_x \pm i  t_y) \vert T_{\pm} \rangle \langle S \vert + \text{H.c.} \right).
	\end{align}
Often, however, the SOI is parametrized by a spin-orbit field $\mathbf{M}_{SO}^d$ in dot $ d \in \lbrace L, R \rbrace$,
	\begin{align}
	\label{eq:spin_orbit_field}
		H_{SO} = \frac{\mu_B}{2} \sum_d \mathbf{M}_{SO}^d \cdot \boldsymbol{\sigma}^d,
	\end{align}
where $\boldsymbol{\sigma}^d$ is a vector containing the Pauli matrices in dot $d$. A general argument shows that the spin-orbit vector $\mathbf{t}_{SO}$ cannot be mapped to the spin-orbit field $\mathbf{M}_{SO}$. On the one hand, the Hamiltonian in Eq.~\eqref{eq:H_SO_spin_orbit_vector} is time reversal invariant while the Zeeman type Hamiltonian~\eqref{eq:spin_orbit_field} with a real spin-orbit field is not. On the other hand, a complex spin-orbit field would make~\eqref{eq:spin_orbit_field} non Hermitian, and hence the two quantities $\mathbf{t}_{SO}$ and $\mathbf{M}_{SO}$ cannot be related. Taking the spin-orbit field to be a function of the crystal momenta, $\mathbf{M}_{SO}^d(\mathbf{k})$, can restore time reversal symmetry but is not an appropriate description for the confined hole spin states considered here. Another quantity is based on an effective approach to first order in the SOI, explicitly taking into account phonons~\cite{Golovach2004},
	\begin{align}
	\label{eq:effective_SO_field}
		\mathbf{M}^d_{SO} = 2 \mathbf{B} \times \boldsymbol{\Omega}^d_{SO},
	\end{align}
where $\mathbf{B}$ is the applied magnetic field, and $\boldsymbol{\Omega}_{SO}$ contains the effect of the SOI in dot $d$. However, with this definition one has $\mathbf{M}_{SO} = 0$ at zero magnetic field, in contrast to Eq.~\eqref{eq:H_SO_spin_orbit_vector} which is independent of the magnetic field. Rather, a Zeeman term~\eqref{eq:spin_orbit_field} with a spin-orbit field of the form~\eqref{eq:effective_SO_field} complements the spin-flip tunneling terms appearing in the spin-orbit vector with additional intra-dot processes stemming from higher orbital contributions. These are expected to be negligible when the orbital energies $\hbar \omega_{x,y}$ are large compared to the spin-orbit parameters (in the present system we find $\vert \mathbf{t}_{SO} \vert \lesssim 500$~neV, $\hbar \omega_{x,y} \sim 1$~meV).

Yet another quantity commonly used in the literature is the spin-orbit length $l_{SO}$. It is shown in Ref.~\cite{Mutter2021} that the spin-conserving and spin-flip tunneling terms in a Ge double quantum dot with cubic Rashba SOI and a locally harmonic double quantum well potential are given by
	\begin{align}
	\label{eq:tunnel_matrix_elements}
	\begin{split}
		& t_C =  \frac{3N\gamma}{4} \left( \hbar \omega_0  + \frac{d^2 m  \omega_0^2}{4} \right),\\
		&  t_{SO}  =\lambda_R   N (S-\gamma)   \frac{d^3 m^3 \omega_0^3}{4} ,
	\end{split}
	\end{align}
where $\hbar \omega_0$ is the in-plane confinement energy, $m$ is the in-plane HH mass, $d$ is the interdot distance, $S$ is the overlap between the left and right dot states, $\gamma = (1- \sqrt{1-S^2})/S$, and $\lambda_R$ is the effective cubic Rashba coefficient with units $[\lambda_R] = $ energy/momentum$^3$. We neglect the effect of excited orbitals and orbital magnetic contributions which are expected to be valid simplifications at the confinement energies and low magnetic fields considered in this paper. Following Ref.~\cite{Stepanenko2012}, we express the spin-flip tunneling element in terms of the spin-conserving tunneling element,
	\begin{align}
	\label{eq:t_so_in_terms_of_tc}
		t_{SO} = \frac{4 t_C}{3} \frac{d}{l_{SO}},
	\end{align}
and use this expression as a working definition of the spin-orbit length,
	\begin{align}
	\label{eq:spin_orbit_length}
		l_{SO} = \frac{1}{\lambda_R} \frac{\gamma}{S- \gamma} \frac{4 \hbar \omega_0 + d^2 m \omega_0^2}{d^2 m^3 \omega_0^3} = \frac{1}{\lambda_R} \frac{1}{\sqrt{1-S^2}} \frac{4 \hbar \omega_0 + d^2 m \omega_0^2}{d^2 m^3 \omega_0^3}.
	\end{align}
Ref.~\cite{Stepanenko2012} considers a DQD with a linear SOI, and we find that the spin-orbit length is rather different for systems with a cubic Rashba SOI. In particular, it depends not only on material parameters via $\lambda_R$ but also on the dot geometry via $\omega_0$, $d$, $S$ and $\gamma$.

This dependence on the dot geometry can be understood from the definition of the spin-orbit length as the distance travelled by a particle before its spin is flipped. Assume for simplicity a particle travelling along $x$ (e.g. the DQD axis) and an $n$th order Rashba type SOI of the form
	\begin{align}
		\label{eq:n_th_order_Rashba}
			H_{SO} =   \alpha_n  (\hbar k_x)^n \sigma_y,
	\end{align}
where the real number $\alpha_n$ is the coupling strength with units energy/momentum$^n$, $k_x$ the crystal momentum along $x$ and $\sigma_y$ a Pauli matrix. We may then work with an effective spin Hamiltonian by averaging over the crystal momentum,
	\begin{align}
			\langle H_{SO} \rangle_{\psi} =   \alpha_n \hbar^n  \langle k_x^n \rangle_{\psi} \sigma_y,
	\end{align}
where the state $\psi$ is only required to satisfy the condition $\langle k_x^n \rangle_{\psi} \neq 0$ to allow for finite spin flip times. In this case the time evolution generated by this Hamiltonian in spin space will cause a spin flip in time $\tau = \pi/(2 \alpha_n \hbar^{n-1} \langle k_x^n \rangle_{\psi} )$. The spin-orbit length along $x$ is then defined to be
	\begin{align}
	\label{eq:SO_length_general}
		l_{SO} := \frac{\langle p_x \rangle_{\psi}}{m} \tau = \frac{\hbar \langle k_x \rangle_{\psi} }{m} \tau = \frac{\pi}{2} \frac{\hbar^{2-n}}{m \alpha_n} \frac{\langle k_x \rangle_{\psi} }{\langle k_x^n \rangle_{\psi} }.
	\end{align}
For $n=1$, the final fraction equals one and the spin-orbit length is a well defined quantity which only depends on material parameters. However, for $n >1$ the expectation values must be evaluated given a specific state of the system. This introduces two problems: Firstly, since the state must be chosen by hand there is a degree of arbitrariness in the definition of the spin-orbit length. Secondly, given a specific setup, the state chosen for computing the expectation value will depend on the system geometry, and hence the spin-orbit length loses its character as a geometry independent quantity as was explicitly shown above for the case of the cubic ($n=3$) Rashba SOI. It is therefore necessary to treat the spin-orbit length with caution for nonlinear SOI as it can possibly no longer be considered a good figure of merit for material comparison.

\section{Effect of the dot geometry on the $g$-factors}
\label{sec:DotGeo}
Heavy-hole (HH) and light-hole (LH) states confined in a germanium heterostructure in the presence of a magnetic field $\mathbf{B}$ are well described by the Hamiltonian
	\begin{align}
	\label{eq:Hamiltonian}
		H = -\frac{\hbar^2}{2m_0} \left[ \left( \gamma_1 + \frac{5}{2} \gamma_s \right) \mathbf{k}^2- 2 \gamma_s \left( \mathbf{k} \cdot \mathbf{J}  \right)^2 \right] + 2 \mu_B \left( \kappa \mathbf{B} \cdot \mathbf{J} + q \sum_{n=1}^3  B_n J_n^3  \right) + V(z) + U(x,y).
	\end{align}
Here, the first term is the Luttinger-Kohn Hamiltonian in spherical approximation featuring the Luttinger parameters $\gamma_1$ and $\gamma_s$, the bare electron mass $m_0$ and the spin $3/2$ matrices $J_n$. The second term describes the effect of a magnetic field in the context of the envelope function approximation, $\kappa$ and $q$ being constants that depend on system specific influences such as strain and material composition~\cite{Marie1999, Winkler2003, Trifonov2021}. We model the heterostructure by an infinite hard wall potential $V(z)$ in the out-of-plane direction, and realize the quantum dot in the $x$-$y$-plane by an elliptical harmonic confinement potential,
	\begin{align}
		U(x,y) = \frac{m}{2}  \left( \omega^2_x x^2 + \omega^2_y y^2 \right).
	\end{align}
Note that the confinement energies $\omega_x$ and $\omega_y$ are defined with respect to the in-plane HH mass $m = m_0/(\gamma_1 + \gamma_s)$.  In the following we assume small magnetic fields in the mT range and neglect orbtial effects arising from the canonical momentum. This is valid for in-plane fields $B_{x,y}$ if $\epsilon_{\parallel} = \hbar e B_{x,y} /2m \ll \Delta = 2 \gamma_s \hbar^2 \pi^2/m_0 d^2$, where $\Delta$ is the HH-LH splitting with the quantum well width $d$. At $d \approx 20$~nm one finds $\Delta \approx 17$~meV and hence $  \epsilon_{\parallel} \ll \Delta$ for $ B_{x,y} \lesssim 1$~T. On the other hand, for out-of-plane fields $B_z$ we require $\epsilon_{\perp} = \hbar e B_z/2m \ll  \text{min} \lbrace  \hbar \omega_x, \hbar \omega_y \rbrace$. For a typical minimal in-plane confinement energy $ \text{min} \lbrace \hbar \omega_x, \hbar \omega_y \rbrace \approx 0.5$~meV, we have  $\epsilon_{\perp} / \text{min} \lbrace \hbar \omega_x,  \hbar \omega_y \rbrace < 0.05$ at $B_z = 20$~mT which is the maximal field applied in this work.

The eigenstates of the Hamiltonian~\eqref{eq:Hamiltonian} in the absence of HH-LH mixing and at zero in-plane magnetic field have the form $\vert \Psi_{h} \rangle  \vert j_z \rangle$ ($\vert \Psi_{l} \rangle  \vert j_z \rangle$) for $j_z =\pm 3/2$ ($j_z =\pm 1/2$), where $j_z$ is the magnetic quantum number of the total angular momentum, and the position states $\vert \Psi_{h/l} \rangle$ are a product of simple harmonic oscillator states in $x$ and $y$ and trigonometric functions satisfying the boundary conditions set by $V(z)$ at $\pm d/2$. We proceed to project the Hamiltonian~\eqref{eq:Hamiltonian} onto the space spanned by the orthonormal orbital ground state basis $\lbrace \vert \Psi^0_{h} \rangle \vert 3/2 \rangle, \vert \Psi^0_{l} \rangle \vert 1/2 \rangle,\vert \Psi^0_{l} \rangle \vert -1/2 \rangle,\vert \Psi^0_{h} \rangle \vert -3/2 \rangle \rbrace$ with
	\begin{align}
	\label{eq:wave_functions}
	\begin{split}
		 &\Psi^0_{h/l} (x,y,z) = \psi_{h/l}(x,y)  \phi(z) , \\
		 &\psi_{h/l}(x,y) = \sqrt{\frac{m_{h/l} \sqrt{\omega_x \omega_y}}{\hbar \pi}} \exp \left( -\frac{m_{h/l} }{2 \hbar} \left[ \omega_x x^2 + \omega_y y^2 \right] \right), \quad \phi(z) = \sqrt{\frac{2}{d}} \cos \left(  \frac{\pi z}{d}  \right) \Theta \left( \frac{d}{2} - \vert z \vert \right),
	\end{split}
	\end{align}
where $\Theta$ is the Heaviside step function, $m_h = m$ and $m_l = \sqrt{ M m}$ with in-plane LH mass $M = m_0/ (\gamma_1 - \gamma_s)$. Note that $m < M$ for in-plane motion. One obtains a $4\times 4$ matrix (labelled by angular momentum) in which the HH and LH spaces are separated in energy by the HH-LH splitting $\Delta $ and can mix due to the intrinsic SOI,
	\begin{align}
	\label{eq:H_LK_matrix}
		\langle H \rangle = \begin{pmatrix}
			\left(3 \kappa + \frac{27}{4}q \right) \mu_B  B_z & \sqrt{3} \left( \kappa + \frac{7}{4}q \right) \mu_1 \mu_B b^*  & -\sqrt{3} \gamma_s \mu_2 \hbar (\omega_x - \omega_y) & \frac{3}{2} q \mu_B b \\
			\sqrt{3} \left( \kappa + \frac{7}{4}q \right) \mu_1 \mu_B b & \epsilon + \left( \kappa + \frac{1}{4}q \right) \mu_B B_z & (5 q + 2 \kappa) \mu_B b^* & -\sqrt{3} \gamma_s \mu_2 \hbar (\omega_x - \omega_y) \\
			-\sqrt{3} \gamma_s \mu_2 \hbar (\omega_x - \omega_y) &  (5 q + 2 \kappa)  \mu_B b & \epsilon - \left( \kappa + \frac{1}{4}q \right) \mu_B B_z & \sqrt{3} \left( \kappa + \frac{7}{4}q \right) \mu_1  \mu_B b^*\\
			\frac{3}{2} q \mu_B b^* & -\sqrt{3} \gamma_s \mu_2 \hbar (\omega_x - \omega_y)  & \sqrt{3} \left( \kappa + \frac{7}{4}q \right) \mu_1 \mu_B b & - \left(3 \kappa + \frac{27}{4}q \right) \mu_B B_z 
		\end{pmatrix},
	\end{align}
where we introduce the quantities
	\begin{align}
		& b= B_x + i B_y, \quad \epsilon = \frac{\hbar (\omega_x + \omega_y)}{2} \left(\sqrt{\frac{m}{M}} -1 \right) + \Delta,  \quad \mu_1 = \langle \psi_h \vert \psi_l \rangle =  \frac{2 (m M)^{1/4} }{\sqrt{m} + \sqrt{ M}}, \quad \mu_2 = \frac{m^{3/2} (m M)^{3/4} }{m_0 (m + \sqrt{m M})^2}.
	\end{align}
Note that any additional part of the Hamiltonian that is linear in the momentum (e.g. a Rashba type SOI) vanishes upon projection onto the space spanned by the symmetric ground state wave functions. Assuming the HH-LH splitting to be the largest relevant energy scale, $\Delta \gg \mu_B \vert \mathbf{B} \vert, \hbar \omega_x, \hbar \omega_y $, we may perform a Schrieffer-Wolff transformation to decouple the LH subspace. We find an effective $2 \times 2$ HH Hamiltonian with a Zeeman term (i.e., a term linear in $\mathbf{B}$) containing the renormalized g-factors
	\begin{subequations}
	\begin{align}
	\label{eq:effective_g_factors}
		&g_x= g_x^0 -\xi_1 \frac{\hbar(\omega_x - \omega_y) }{\hbar(\omega_x + \omega_y) - \xi_2 \Delta}, \\
		&g_y = g_y^0 + \xi_1 \frac{\hbar(\omega_x - \omega_y) }{\hbar(\omega_x + \omega_y) - \xi_2 \Delta}, \\
		& g_z =  g_z^0  -  \xi_1 \left[ \frac{\hbar(\omega_x - \omega_y) }{\hbar(\omega_x + \omega_y) - \xi_2 \Delta} \right]^2,
	\end{align}
	\end{subequations}
where $g_x^0 = g_y^0 = 3q$, $g_z^0 = 6 \kappa + 27 q/2$ are the effective zeroth order values and
	\begin{align}
		\xi_1 = \left(12 \kappa + 21q \right)  \frac{\sqrt{\gamma_1^2- \gamma_s^2}  }{  \gamma_1 + \sqrt{\gamma_1^2 - \gamma_s^2} }, \quad \xi_2 = \frac{2 \sqrt{\gamma_1+ \gamma_s}}{\sqrt{\gamma_1+ \gamma_s} - \sqrt{\gamma_1 - \gamma_s}}
	\end{align}
are material specific constants, $\xi_1 \approx 20.3$ and $\xi_2 \approx 6.0$ in (bulk) germanium. While $g_z$ is always reduced by the HH-LH mixing, the sign of the correction to $g_x$ and $g_y$ depends on the relative in-plane confinement strength. In particular, $g_x \neq g_y$ due to the broken in-plane symmetry. It is therefore possible that one in-plane g-factor becomes negative. This effect can be observed even in a perturbative approach because the zeroth order in-plane g-factors in germanium are small, $g_x^0 = g_y^0 = 3q \approx 0.2$. In contrast, we find that the correction to the out-of-plane g-factor is $\vert \delta g_z \vert \lesssim 10^{-2}$ for the values considered here and hence negligible. The smallness of the correction to the out-of-plane g-factor compared to the correction to the in-plane g-factors can be understood as follows: The HH and LH subspaces are mixed by the effect of elliptical confinement ($\sim [\omega_x - \omega_y]$) and the in-plane magnetic fields, and the corrections to the matrix elements in the effective HH Hamiltonian are given by a product of these effects. The effective in-plane g-factor corrections then effectively contain only one of the two small mixing terms ($\delta g_{x,y} \sim [\omega_x - \omega_y]$)) because one factor contains the in-plane magnetic field, while the out-of-plane g-factor correction contains two small factors ($\delta g_z \sim [\omega_x - \omega_y]^2$).

\end{document}